\documentclass[english]{article}
\usepackage{amsmath,bm,amssymb,soul,mathtools,mathptmx}
\usepackage[T1]{fontenc}
\usepackage[letterpaper]{geometry}
\usepackage{float}
\usepackage{amsmath}
\usepackage{graphicx}
\usepackage{color}
\usepackage{epstopdf}

\usepackage[colorlinks, hypertexnames=false]{hyperref}
\hypersetup{linkcolor=blue, citecolor=red} % set color for references
\usepackage{adjustbox}
\usepackage{bm}
\usepackage{multirow}
\usepackage{array}
\usepackage{mathrsfs}
\usepackage{amsmath, amssymb, amsthm}

\usepackage{subfigure}
\usepackage{tikz}
\usepackage{algorithm}
\usepackage{algpseudocode}

\usepackage{amsfonts} 
\usepackage{color}
\usepackage{adjustbox}
\definecolor{black}{rgb}{0,0,0}

\definecolor{red}{rgb}{1,0,0}

\definecolor{blue}{rgb}{0,0,1}

\usepackage{multirow}

\DeclareOldFontCommand{\rm}{\normalfont\rmfamily}{\mathrm}
\DeclareOldFontCommand{\sf}{\normalfont\sffamily}{\mathsf}
\DeclareOldFontCommand{\tt}{\normalfont\ttfamily}{\mathtt}
\DeclareOldFontCommand{\bf}{\normalfont\bfseries}{\mathbf}
\DeclareOldFontCommand{\it}{\normalfont\itshape}{\mathit}
\DeclareOldFontCommand{\sl}{\normalfont\slshape}{\@nomath\sl}
\DeclareOldFontCommand{\sc}{\normalfont\scshape}{\@nomath\sc}
\DeclareRobustCommand*\cal{\@fontswitch\relax\mathcal}
\DeclareRobustCommand*\mit{\@fontswitch\relax\mathnormal}

\makeatother
\usepackage{mathtools}
\usepackage{esint}
\usepackage{cleveref}
\usepackage{natbib}
\usepackage{babel}
\usepackage{authblk}

\def\topline{\hline\hline\vrule height 10pt depth4pt width0pt\relax}
\def\midline{\hline\vrule height 10pt width0pt\relax}
\def\botline{\hline}
\def\acknowledgments{\paragraph*{Acknowledgments.}}

\def\datastatement{\paragraph*{Data availability statement.}}

\title{\textbf{A Closed-Form Nonlinear Data Assimilation Algorithm for Multi-Layer Flow Fields}}

\author[1]{Zhongrui Wang}
\author[1,*]{Nan Chen}
\author[2]{Di Qi}
\affil[1]{Department of Mathematics, University of Wisconsin–Madison, Madison, WI 53706, USA}
\affil[2]{Department of Mathematics, Purdue University, West Lafayette, IN 47907, USA}
\affil[*]{\small{Corresponding author: Nan Chen, chennan@math.wisc.edu}}

\begin{document}
	\maketitle
	\begin{abstract}
State estimation in multi-layer turbulent flow fields with only a single layer of partial observation remains a challenging yet practically important task. Applications include inferring the state of the deep ocean by exploiting surface observations. Directly implementing an ensemble Kalman filter based on the full forecast model is usually expensive. One widely used method in practice projects the information of the observed layer to other layers via linear regression. However, when nonlinearity in the highly turbulent flow field becomes dominant, the regression solution will suffer from large uncertainty errors. In this paper, we develop a multi-step nonlinear data assimilation method. A sequence of nonlinear assimilation steps is applied from layer to layer recurrently. Fundamentally different from the traditional linear regression approaches, a conditional Gaussian nonlinear system is adopted as the approximate forecast model to characterize the nonlinear dependence between adjacent layers. The estimated posterior is a Gaussian mixture, which can be highly non-Gaussian. Therefore, the multi-step nonlinear data assimilation method can capture strongly turbulent features, especially intermittency and extreme events, and better quantify the inherent uncertainty. Another notable advantage of the multi-step data assimilation method is that the posterior distribution can be solved using closed-form formulae under the conditional Gaussian framework. Applications to the two-layer quasi-geostrophic system with Lagrangian data assimilation show that the multi-step method outperforms the one-step method with linear stochastic flow models, especially as the tracer number and ensemble size increase.
\end{abstract}

%%%%%%%%%%%%%%%%%%%%%%%%%%%%%%%%%%%%%%%%%%%%%%%%%%%%%%%%%%%%%%%%%%%%%
% MAIN BODY OF PAPER
%%%%%%%%%%%%%%%%%%%%%%%%%%%%%%%%%%%%%%%%%%%%%%%%%%%%%%%%%%%%%%%%%%%%%
%
\section{Introduction}
Data assimilation has been widely applied in state estimation in the atmosphere and ocean \citep{ghil_data_1991, kalnay_atmospheric_2002, griffa_lagrangian_2007, kalnay_ncepncar_2011}. Many practical applications require estimating the states of a multi-layer flow field using observations from only a single layer. Examples include inferring the state of the deep ocean by exploiting surface observations \citep{molcard_lagrangian_2005} and estimating upper atmospheric conditions using near-surface observations \citep{mou2023combining}. These problems are particularly challenging since information gets lost as it propagates across layers due to its turbulent nature and the associated uncertainties. Addressing such challenges requires methods capable of accurately capturing information transferred across layers in highly turbulent systems.

One natural way to assimilate a multi-layer flow field is to implement an ensemble Kalman filter (EnKF) directly \citep{evensen2022data, carrassi2018data,lee2017preventing} based on the full forecast model. The advantage of such a method is that the forecast model accounts for the nonlinear dependence between layers characterized by the forecast model, allowing an effective information transfer across adjacent layers. Yet, the main difficulty in such a method lies in its computational cost since running the full forecast model is usually expensive, let alone generating an ensemble of simulations. One widely used alternative in practice is to project the information of the observed layer to other layers via linear regression \citep{molcard_lagrangian_2005}. However, due to the highly turbulent nature of the underlying flow field, the accuracy of the state estimation via linear regression may suffer from large uncertainty.

This paper aims to develop an efficient nonlinear data assimilation algorithm that effectively propagates information from the observed to unobserved layers. Our primary focus is on a widely relevant oceanographic problem: recovering ocean velocity fields from surface tracer observations, although the method holds significant potential for a wide range of atmospheric and climate science issues. Tracers are Lagrangian observations that move with the flow. They are particularly useful for the state estimation of ocean flow fields, where the deployment of fixed Eulerian observation stations is challenging \citep{griffa_lagrangian_2007, molcard_assimilation_2003, gould_argo_2004}. Examples include ocean surface drifters \citep{centurioni_global_2017}, deep-ocean profiling floats \citep{gould_argo_2004}, sea ice floes \citep{mu_arctic-wide_2018, chen_efficient_2022}, and ballons \citep{businger_balloons_1996}. Beyond the multi-layer flow structures, data assimilation exploiting Lagrangian tracers, known as Lagrangian data assimilation \citep{mariano_lagrangian_2002,ide_lagrangian_2002, kuznetsov_method_2003,molcard_assimilation_2003, molcard_lagrangian_2005}, introduces another unique challenge: the observational process is intrinsically nonlinear. The nonlinear observational process adds further complexity to assimilating multi-layer flow fields, as the nonlinearities in both the observational process and the interdependence between layers need to be handled appropriately.

Early practices of Lagrangian data assimilation ignore the nonlinearity and treat the Lagrangian observations as Eulerian \citep{carter_assimilation_1989, ishikawa_successive_1996}. \cite{ide_lagrangian_2002} addresses the nonlinearity by augmenting the flow model with a tracer dynamical model, with the posterior solved by the extended Kalman filter (EKF). \cite{molcard_assimilation_2003} explicitly considers the variational derivative of the observation operator based on the optimal interpolation (OI) method. Despite promising results in experiments with quasi-geostrophic and primitive equation models \citep{molcard_assimilation_2003, ozgokmen_assimilation_2003}, it has been proven that an accurate linear approximation of the observation operator is nearly unreachable \citep{piterbarg_optimal_2008}. \cite{apte_bayesian_2008} applied the EnKF in Lagrangian data assimilation followed by investigating the nonlinear effects on EnKF \citep{apte_impact_2013}. In addition to its nonlinear nature, Lagrangian data assimilation often involves high-dimensional systems, making advanced numerical methods, such as the particle filter (PF) \citep{van_leeuwen_particle_2019} and the Markov chain Monte Carlo (MCMC) as smoother \citep{apte_bayesian_2008}, computationally expensive. Recent advancements in higher-dimensional Lagrangian data assimilation include the development of hybrid strategies \citep{slivinski_hybrid_2015} and localized methods, such as the use of the local ensemble transform Kalman filter (LETKF) in the Geophysical Fluid Dynamics Laboratory (GFDL) Modular Ocean Model \citep{sun_lagrangian_2019} and its application to a sea ice model for assimilating sea ice floe trajectories \citep{chen_efficient_2022}.

\cite{chen_information_2014} proposed a theoretical framework for Lagrangian data assimilation that rigorously allows the nonlinear observation process. The theory was later put into practice \citep{chen_uncertainty_2023}, which involves adopting linear stochastic models (LSMs) as a reduced-order forecast model in the spectral space to reduce computational cost and avoid interpolation errors and extra costs of coordinate transformation. Although the nonlinear Lagrangian data assimilation framework proposed by \cite{chen_information_2014} accounts for the nonlinearity in observations, it is restricted to utilizing linear forecast models for flow variables. Therefore, corrections from the data-containing layer can only linearly propagate across layers. An example of linear projection of corrections based on layer correlations is given by \cite{molcard_lagrangian_2005}. Linear projection methods can be inaccurate when the flow is highly turbulent with deep layers. It is thus crucial to develop a computationally efficient approach that can handle nonlinearity not only in the observational process but also in the flow model, particularly when the nonlinear dependence between variables across different layers is predominant. How to effectively propagate corrections across layers to address flow nonlinearity remains an open question in both Lagrangian and Eulerian data assimilation.

In this paper, a multi-step data assimilation scheme for multi-layer flow fields is developed that handles nonlinearity in both the observational process and the underlying flow field across different layers. The scheme begins by exploiting tracer observations to update the data-containing layer using the nonlinear Lagrangian data assimilation scheme in \cite{chen_uncertainty_2023}. In the subsequent step, samples are drawn from the posterior of the previous assimilated layer and treated as pseudo-observations to update the next layer. A conditional Gaussian nonlinear system is systematically adopted as the approximate forecast model to characterize the nonlinear dependence between adjacent layers, which aims to capture the features of the strong turbulence in the underlying flow and facilitate uncertainty quantification. The resulting posterior is a Gaussian mixture distribution as in \cite{majda2014blended}. The procedure is then repeated sequentially for each layer until all layers are assimilated. The method has several desirable features. First, the Gaussian mixture posterior is a non-Gaussian distribution that has the potential to better characterize intermittency and extreme events appearing in the underlying turbulent flow field. Second, fundamentally different from linear regression, the conditional Gaussian nonlinear forecast model explicitly accounts for nonlinear layer dependencies. Third, the conditional Gaussian nonlinear forecast model allows us to derive a closed-form analytic solution for posterior distribution. The analytic formulae make the solver more robust and accurate.

The rest of the paper is organized as follows. Section 2 describes the multi-step data assimilation for a multi-layer flow field with surface observations. A one-step method with linear stochastic flow models is also described for comparison. Section 3 applies the data assimilation methods to the two-layer quasi-geostrophic system. Section 4 presents the data assimilation results, sensitivities to parameters, and computational cost analysis. Section 5 gives conclusions and discussions.

\section{Data assimilation methods for multi-layer flows}\label{sec:da_multilayer}
The multi-step data assimilation strategy for multi-layer flow fields is presented in this section. First, a general formulation of multi-layer flow fields with only surface observation is described. Then, we introduce the conditional Gaussian nonlinear system, which allows closed analytic formulae for solving posterior statistics and can be integrated seamlessly with the multi-step data assimilation algorithm. We derive the multi-step data assimilation scheme based on a rigorous formulation by applying the proposed nonlinear data assimilation algorithm sequentially to each layer from top to bottom and adopting an efficient ensemble approximation. Note that a one-step data assimilation method with linear stochastic flow models is introduced for comparison.

\subsection{A multi-layer flow system with surface Lagrangian observations}
A multi-layer flow field together with Lagrangian tracer trajectory observations constrained on the surface layer can be described in the general form:
\begin{subequations}\label{eq:multiflow}
\begin{align}
\frac{\mathrm{d}\mathbf{x}_\ell}{\mathrm{d}t} &= \mathbf{v}_1(\mathbf{x}_\ell, t) + \bm{\mathsf{\Sigma}}_{\mathbf{x}} \dot{\mathbf{W}}_\ell, \quad \ \ell = 1, \ldots, L  \label{eq:multilayera} \\
\frac{\mathrm{d}\mathbf{v}}{\mathrm{d}t} &= (\bm{\mathsf{L}}+\bm{\mathsf{D}})\mathbf{v} + \mathbf{B}(\mathbf{v}, \mathbf{v}) + \mathbf{F} + \bm{\mathsf{\Sigma}}_{\mathbf{v}} \dot{\mathbf{W}}_{\mathbf{v}},\label{eq:multiflowb}
\end{align}
\end{subequations}
where $\mathbf{x}_\ell = (x_\ell, y_\ell)^{\mathrm{T}}$ is the observed displacement of the $\ell$th tracer, and $\mathbf{v}=(\ldots,\mathbf{v}_i,\ldots)^{\mathrm{T}},i=1,\ldots,I$ is the unobserved flow velocity field that consists of planar velocities $\mathbf{v}_i=(u_i,v_i)$ of $I$ layers. The tracers are assumed to be massless so that the velocity of each tracer is identical to the surface layer's velocity $\mathbf{v}_1$. In the general formulation of the flow equation, $\bm{\mathsf{L}}$ and $\bm{\mathsf{D}}$ are linear operators representing dispersion and dissipation effects, respectively. $\mathbf{B}(\mathbf{v}, \mathbf{v})$ is a quadratic form that contains nonlinear effects. $\mathbf{F}$ is a deterministic forcing term. The randomness is introduced by $\bm{\mathsf{\Sigma}} \dot{\mathbf{W}}$, where $\dot{\mathbf{W}}$ is the derivative of Wiener process, also known as Gaussian white noise, and $\bm{\mathsf{\Sigma}}$ is the noise strength matrix. The abstract flow formulation (\ref{eq:multiflowb}) models a rich class of turbulent dynamical systems, enabling precise analysis in a list of quantitative and qualitative models \cite{majda_prototype_2016} and many applications \cite{majda2018strategies,qi2023high,qi2023random}.

The massless passive tracer equation (\ref{eq:multilayera}) provides a clean yet comprehensive formulation to investigate the fundamental challenges in Lagrangian data assimilation. Specifically, the tracer velocity $\mathbf{v}_1(\mathbf{x}_\ell, t)$ has a nonlinear dependence on the tracer displacement $\mathbf{x}_\ell$, making the observation process nonlinear. %This nonlinearity arises even without requiring a nonlinear flow model. 
\cite{apte_bayesian_2008} shows that a tracer driven by the linearized shallow water model can already have a nonlinear trajectory. 
On top of that, the flow model (\ref{eq:multiflowb}) introduces additional nonlinearity from its coupling term $\mathbf{B}\left(\mathbf{v},\mathbf{v}\right)$. This intrinsic nonlinearity of Lagrangian data assimilation makes the analysis difficult in theory and the optimal solution almost intractable in practice \citep{ide_lagrangian_2002,molcard_assimilation_2003, chen_information_2014}.

\subsection{The conditional Gaussian  nonlinear system}
The conditional  Gaussian nonlinear system (CGNS) is a rich class of nonlinear systems that has the following structure \citep{liptser2013statistics, chen_conditional_2018, chen_conditional_2022}:
\begin{subequations}\label{eq:cgns}
\begin{align}
        \frac{\mathrm{d} \mathbf{u}_1}{\mathrm{d}t} = \mathbf{A}_0(\mathbf{u}_1,t) + \bm{\mathsf{A}}_1(\mathbf{u}_1,t)\mathbf{u}_2 + \bm{\mathsf{\Sigma}}_1(\mathbf{u}_1,t) \dot{\mathbf{W}}_1, \\
         \frac{\mathrm{d} \mathbf{u}_2}{\mathrm{d}t} = \mathbf{a}_0(\mathbf{u}_1,t) + \bm{\mathsf{a}}_1(\mathbf{u}_1,t)\mathbf{u}_2 + \bm{\mathsf{\Sigma}}_2(\mathbf{u}_1,t)  \dot{\mathbf{W}}_2,
\end{align}
\end{subequations}
where $\mathbf{u}_1\in \mathbb{C}^{N_1}$ and $\mathbf{u}_2\in \mathbb{C}^{N_2}$ are vectors of complex state variables. $\mathbf{A}_0$ and $\mathbf{a}_0$ are vectors. $\bm{\mathsf{A}}_1$ and $\bm{\mathsf{a}}_1$ are matrices. $\bm{\mathsf{\Sigma}}_1\dot{\mathbf{W}}_1$ and $\bm{\mathsf{\Sigma}}_2\dot{\mathbf{W}}_2$ are independent white noises multiplied by noise strength matrices. A rich class of turbulent systems belongs to the CGNS family, including the noisy Lorenz 63 system \citep{lorenz_deterministic_1963}, the Boussinesq equation, and the rotating shallow water equations, to name a few. Many other systems can systematically be approximated by conditional Gaussian statistical models, which greatly enriches the application of CGNS in fluid dynamics. More examples and applications can be found in \cite{chen_conditional_2018,chen2024physics}.

One notable feature of  the CGNS is that $\mathbf{A}_0$,  $\mathbf{a}_0$, $\bm{\mathsf{A}}_1$, $\bm{\mathsf{a}}_1$, $\bm{\mathsf{\Sigma}}_1$ and $\bm{\mathsf{\Sigma}}_2$ are nonlinear functions of $\mathbf{u}_1$.  As a result, the whole coupled system can be highly nonlinear, and the marginal distributions of $\mathbf{u}_1$ and $\mathbf{u}_2$ can be strongly non-Gaussian. Another important feature, that there are only linear terms of $\mathbf{u}_2$ in (\ref{eq:cgns}), brings a crucial feature of CGNS --- the conditional Gaussian property. That is, once a trajectory of $\mathbf{u}_1(s)$ for $s\leq t$ is given, $\mathbf{u}_2$ conditioned on $\mathbf{u}_1(s)$ becomes a linear Gaussian process. Therefore, the conditional distribution is Gaussian
\begin{equation} \label{eq:3}
  p(\mathbf{u}_2(t) | \mathbf{u}_1(s \leq t)) \sim \mathcal{N} \big(\boldsymbol{\mu}_2(t), \bm{\mathsf{R}}_2(t) \big),
\end{equation}
with mean $\boldsymbol{\mu}_2(t)$ and covariance $\bm{\mathsf{R}}_2(t)$ solvable through closed analytic formulae \citep{liptser2013statistics}
\begin{subequations}\label{eq:cg_meanvar}
\begin{align}
    \frac{\mathrm{d} \boldsymbol{\mu}_2}{\mathrm{d}t} &= (\mathbf{a}_0+ \bm{\mathsf{a}}_1\boldsymbol{\mu}_2) + \bm{\mathsf{R}}_2\bm{\mathsf{A}}_1^*(\bm{\mathsf{\Sigma}}_1\bm{\mathsf{\Sigma}}_1^*)^{-1}
        \left(\frac{\mathrm{d}\mathbf{u}_1}{\mathrm{d}t}-(\mathbf{A}_0+\bm{\mathsf{A}}_1\boldsymbol{\mu}_2)\right), \label{eq:mean_cg}\\
        \frac{\mathrm{d} \bm{\mathsf{R}}_2}{\mathrm{d}t} &= \bm{\mathsf{a}}_1\bm{\mathsf{R}}_2 +\bm{\mathsf{R}}_2\bm{\mathsf{a}}_1^*+\bm{\mathsf{\Sigma}}_2\bm{\mathsf{\Sigma}}_2^*
    - (\bm{\mathsf{R}}_2\bm{\mathsf{A}}_1^*)(\bm{\mathsf{\Sigma}}_1\bm{\mathsf{\Sigma}}_1^*)^{-1}(\bm{\mathsf{A}}_1\bm{\mathsf{R}}_2
    ). \label{eq:var}
\end{align}
\end{subequations}
In data assimilation, $\mathbf{u}_1$ is considered as the observation with continuous-in-time trajectories, while $\mathbf{u}_2$ is the unobserved state to be filtered, and $ p\big(\mathbf{u}_2(t) | \mathbf{u}_1(s \leq t) \big)$ is the posterior distribution we aim for. With the analytic formulae (\ref{eq:cg_meanvar}), $\boldsymbol{\mu}_2(t)$ and  $\bm{\mathsf{R}}_2(t)$ can be solved exactly. Thus, the CGNS is useful for state estimation, parameter estimation, and uncertainty quantification, especially in complex high-dimensional systems. We refer to data assimilation based on solving (\ref{eq:cg_meanvar}) as conditional Gaussian data assimilation (CGDA) hereafter.

\subsection{Multi-step data assimilation based on the multi-layer flow and CGDA}\label{sec:2c}
Comparing the flow-observation system (\ref{eq:multiflow}) to the CGNS (\ref{eq:cgns}), the flow equations (\ref{eq:multiflow}) can be fitted into the CGNS framework by dropping the quadratic term $\mathbf{B}(\mathbf{v}, \mathbf{v})$ (and assuming constant noise strengths). However, this may lead to severe model simplification in most real-world applications, as nonlinearities often contribute a dominate part of the flow model's dynamics. A multi-step data assimilation method is proposed here, bypassing oversimplification and preserving parts of the flow model's quadratic nonlinearities.

Instead of treating the coupled multi-layer flow-observation system altogether, we perform data assimilation layer by layer, from top to bottom (i.e., from surface to deep ocean). First, assimilate tracer observations to recover the surface-layer flow.  Then, generate samples from the surface-layer flow posterior. These samples are treated as pseudo-observations to recover the lower-layer flow in the second data assimilation step. The new posterior is sampled and this procedure is sequentially applied proceeding to the following lower layer until the bottom layer is recovered. At each data assimilation step, the upper layer is considered as the observation, and the lower layer is to be recovered.

The multi-step data assimilation procedure described above is formulated as follows. Without loss of generality, consider a two-layer flow field with surface observations. We are interested in the conditional probability density function (PDF) (which is the posterior distribution in data assimilation),
\begin{equation}\label{eq:post}
    p\big(\mathbf{v}(t)|\{{\mathbf{x}(s)\}}_{s\leq t}\big),
\end{equation}
where $\mathbf{v}(t)=(\mathbf{v}_1(t), \mathbf{v}_2(t))^{\mathrm{T}}$ are the planar velocities at time $t$, with subscripts indicating the respective layers. $\{\mathbf{x}(s)\}_{s\leq t}$ is one realization of the tracer trajectories up to time $t$. The upper-layer (surface-layer) posterior is straightforward to obtain, as the surface tracers are directly driven by the surface flow. We focus on the lower-layer posterior, which can be written in the marginalized form
\begin{equation}\label{eq:margin}
\begin{aligned}
       p\big({\mathbf{v}}_2(t)|\{{\mathbf{x}(s)\}}_{s\leq t}\big) &= \int_{\Omega_{\mathbf{v}_1(t)}}p\big({\mathbf{v}}_1(t),{\mathbf{v}}_2(t)|\{{\mathbf{x}(s)\}}_{s\leq t}\big)d{\mathbf{v}_1} \\
       &= \int_{\Omega_{\mathbf{v}_1(t)}}p\big({\mathbf{v}}_2(t)|{\mathbf{v}}_1(t),\{{\mathbf{x}(s)\}}_{s\leq t}\big) p\big({\mathbf{v}}_1(t)|\{{\mathbf{x}(s)\}}_{s\leq t}\big) d{\mathbf{v}_1},
\end{aligned}
\end{equation}
where the conditional probabilities $ p\big({\mathbf{v}}_1(t)|\{{\mathbf{x}(s)\}}_{s\leq t}\big) $ and $p\big({\mathbf{v}}_2(t)|{\mathbf{v}}_1(t),\{{\mathbf{x}(s)\}}_{s\leq t}\big)$ can be given by the corresponding upper-layer and lower-layer posteriors in data assimilation. However, one needs to be very cautious here, as there is a nuance between conditional probability and posterior that could introduce errors. This will be explained in more detail at the end of this subsection. %For now, let's go with this idea and proceed with the algorithm.

Next, we approximate the integral by sampling from the upper-layer posterior and use these samples as pseudo observations to get the lower-layer posterior. An ideal sampling strategy considers the temporal correlations with dynamical consistency. The approximated lower-layer posterior is given by the empirical average of the finite samples
\begin{equation}\label{eq:montecarlo}
     p\big({\mathbf{v}}_2(t)|\{{\mathbf{x}(s)\}}_{s\leq t}\big)  \approx \frac{1}{N_s}\sum_{n=1}^{N_s} p\big({\mathbf{v}}_2(t)|{\mathbf{v}}_1^{(n)}(t),\{{\mathbf{x}(s)\}}_{s\leq t}\big),
\end{equation}
where ${\mathbf{v}}_1^{(n)}$ is the $n$th posterior sample out of total $N_s$ samples.

\begin{figure}[h]
 \centerline{\includegraphics[width=39pc]{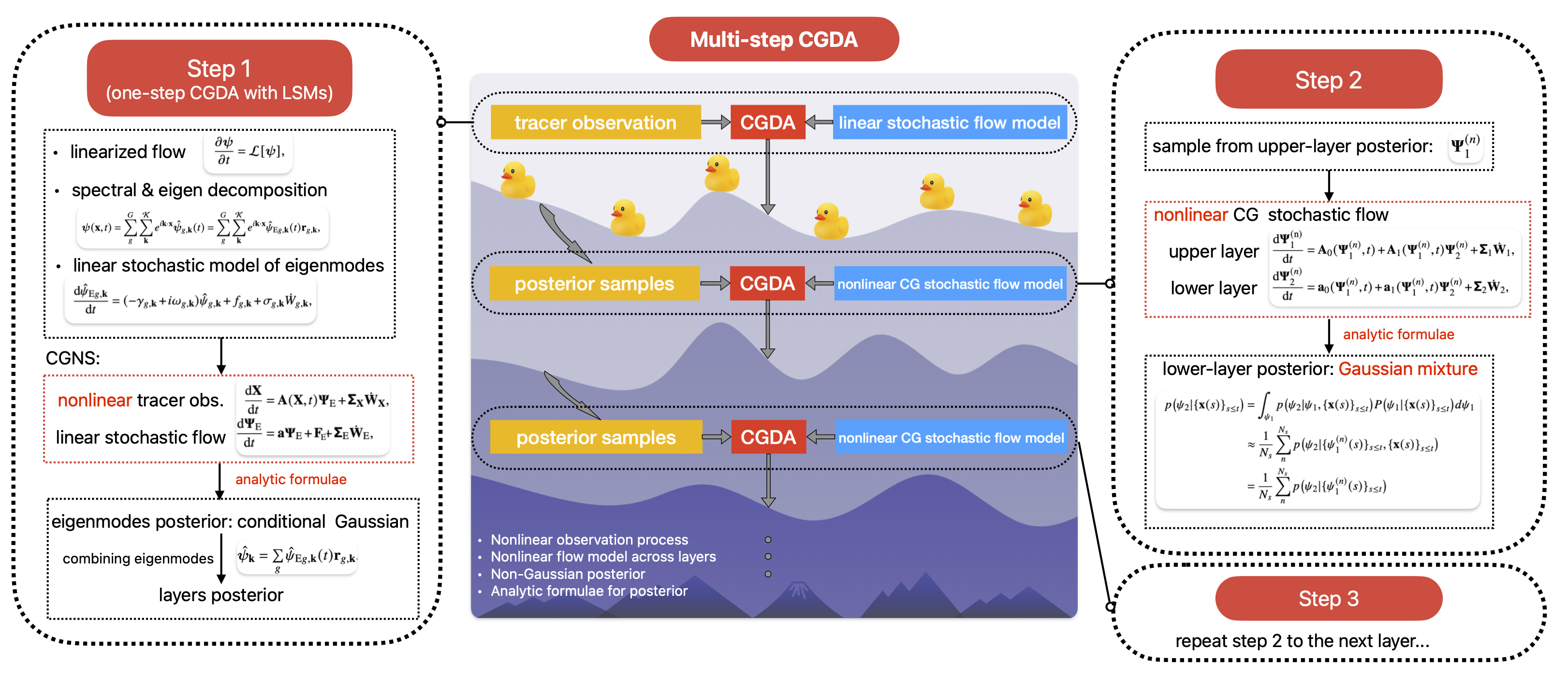}} 
 \caption{An overview of the multi-step CGDA scheme.}
 \label{fig:flowchart}
\end{figure}

Specifically, to apply CGDA in the multi-step data assimilation, we adopt a linear stochastic model to approximate the flow field, and use CGDA to obtain the upper-layer posterior $ p\big({\mathbf{v}}_1(t)|\{{\mathbf{x}(s)\}}_{s\leq t}\big) $ during the first data assimilation step. More details of the first step are given in section \ref{sec:lsmda} and further illustrated in section \ref{sec:3b} with an application to the two-layer quasi-geostrophic system. As the tracer observations appear in the upper layer, they play a much more important role in helping estimate the state in this layer than the subsequent layers. This allows the use of a simplified forecast system based on surrogate linear stochastic dynamics. Next, we sample trajectories $\{\mathbf{v}_1(s)\}_{s\leq t}$ from the upper-layer posterior. An optimal strategy for sampling trajectories of the unobserved variable, given one realization of the observed variable, exists in the CGNS framework \citep{chen_efficient_2020}. The lower-layer posterior is, therefore, approximated by
\begin{equation}\label{eq:cgposterior}
\begin{aligned}
        p\big({\mathbf{v}}_2(t)|{\mathbf{v}}_1^{(n)}(t),\{{\mathbf{x}(s)\}}_{s\leq t}\big) &\approx     p\big({\mathbf{v}}_2(t)|\{{\mathbf{v}}_1^{(n)}(s)\}_{s\leq t},\{{\mathbf{x}(s)\}}_{s\leq t}\big)  \\
        &=p\big({\mathbf{v}}_2(t)|\{{\mathbf{v}}_1^{(n)}(s)\}_{s\leq t}\big).
\end{aligned}
\end{equation}
The condition on tracers can be omitted because tracers do not directly affect the lower-layer flow once the upper-layer flow information is completely given. During the second data assimilation step, we approximate the quadratic terms of $\mathbf{v}_2$ by nonlinear functions of $\mathbf{v}_1$ and conditionally linear functions of $\mathbf{v}_2$ to fit a conditional Gaussian model, and perform CGDA to solve $p\big({\mathbf{v}}_2(t)|\{{\mathbf{v}}_1^{(n)}(s)\}_{s\leq t}\big)$ (see section \ref{sec:3b} for more details). In this way, the nonlinearity of the upper layer is preserved at the second data assimilation step. The resulting lower-layer posterior distribution is a mixture of Gaussians. We refer to the multi-step data assimilation based on CGDA as multi-step CGDA hereafter. An overview of the multi-step CGDA is shown in Fig. \ref{fig:flowchart}.

Notice that a few approximations are introduced in the above deviations that may introduce errors. First, we use an approximate linear flow model to recover the upper-layer flow from tracers and a conditional Gaussian flow model to recover the lower-layer flow. Model errors come in as some nonlinearities are dropped. Second, there are sampling errors in the Monte-Carlo marginalization. In (\ref{eq:margin}) the marginalization is taken over $\mathbf{v}_1(t)$ at one time instance, but in practice, the marginalization is taken over the entire filtered trajectory $\{\mathbf{v}_1(s)\}_{s\leq t}$. This makes the marginalization more difficult because the space of  $\mathbf{v}_1(s), s\leq t$ is very high dimensional. Last, as mentioned earlier, a nuance between conditional probability and posterior can introduce errors from (\ref{eq:margin}) to (\ref{eq:cgposterior}).  According to the standard definition of conditional probability, ${\mathbf{v}}_1$ — on which $\mathbf{v}_2$ is conditioned in $p\big({\mathbf{v}}_2(t)|{\mathbf{v}}_1(t),\{{\mathbf{x}(s)\}}_{s\leq t}\big)$ — should be one realization without uncertainty. However, uncertainty is inherited by sampling from the posterior as pseudo-observations. Therefore, (\ref{eq:margin}) is only an approximation when applying it to assimilating the lower layers.

\subsection{Details on the one-step CGDA with linear stochastic flow models}\label{sec:lsmda}
Here, we provide more details on using a linear stochastic flow model to perform CGDA, following the approach described by \cite{chen_uncertainty_2023}. This will serve as the first step in multi-step CGDA to recover the uppermost layer. Then, the conditional Gaussian nonlinear systems will be adopted as nonlinear forecast models to infer the flow field in the lower layers recursively. See the above subsection for more details. Note that this method can be also used as a stand-alone data assimilation scheme, where the recovery of the lower layers is equivalent to using a linear regression. The performance of such a method will be compared to the multi-step CGDA to explain the benefits of preserving additional quadratic nonlinearities of the flow model in the latter method. It involves a full linearization of the multi-layer flow system, a spectral decomposition, and an eigendecomposition of the linearized flow model and developing reduced-order linear stochastic models (LSMs) for the eigenmodes. The tracer equation (\ref{eq:multilayera}) in conjunction with LSMs for the flow form a CGNS, allowing CGDA to be directly performed. For the simplicity of the statement, consider an incompressible flow field.

First, we drop the nonlinear terms of $\mathbf{v}$ and stochastic terms in (\ref{eq:multiflowb}) to obtain a linearized flow model. Then, we can rewrite the linearized flow equation in terms of the stream function $\boldsymbol{\psi}=({\ldots,\psi_i,\ldots})^{\mathrm{T}}$. The relation between the $i$th layer's incompressible flow velocity and the corresponding stream function is
\begin{equation} \label{eq:v_psi}
    \mathbf{v}_i =\left(\frac{\partial\psi_i}{\partial y}, -\frac{\partial\psi_i}{\partial x}\right)^{\mathrm{T}}.
\end{equation}
This rewriting reduces the number of variables and equations. The linearized stream function equations have the general form of
 \begin{equation}\label{eq:psipde}
     \frac{\partial \boldsymbol{\psi}}{\partial t}=\mathcal{L}[\boldsymbol{\psi}],
 \end{equation}
 where $\mathcal{L}$ is a linear operator following the linear part of (\ref{eq:multiflowb}) (see explicit example next in section \ref{sec:3b}).

 One way to solve such a linear partial differential equations (PDEs) system (\ref{eq:psipde}) is to apply the Fourier transform, in the $\mathbf{x}$ variable, to both sides of the equations \citep{olver_linear_2014}. This will give us a system of ordinary differential equations (ODEs) for the Fourier coefficients $\hat{\boldsymbol{\psi}}_\mathbf{k}(t)$: \begin{equation}\label{eq:psiode}
     \frac{\mathrm{d} \hat{\boldsymbol{\psi}}_\mathbf{k}}{\mathrm{d} t}=\hat{\bm{\mathsf{L}}}_\mathbf{k}\hat{\boldsymbol{\psi}}_\mathbf{k},
 \end{equation}
where the wavenumber $\mathbf{k}=(k_x, k_y) \in  \mathbb{Z}^2$. $\hat{\bm{\mathsf{L}}}_\mathbf{k}$ is a matrix related to the linear operator $\mathcal{L}$. Recall that for a first-order linear ODE system, the general solution is a linear superposition of pure exponentials:
\begin{equation}
\hat{\boldsymbol{\psi}}_\mathbf{k}(t)=\sum_g^G{ e^{\lambda_{g,\mathbf{k}} t}\mathbf{r}_{g,\mathbf{k}}}, \quad g=1,\ldots,G
\end{equation}
where $\{\lambda_{g,\mathbf{k}}\}$ and $\{\mathbf{r}_{g,\mathbf{k}}\}$ are the eigenvalues and eigenvectors of $\hat{\bm{\mathsf{L}}}_\mathbf{k}$, respectively. Let ${ \hat{\psi}_{\mathrm{E} g,\mathbf{k}}(t)}=e^{\lambda_{g,\mathbf{k}} t}$ with $\hat{\psi}_{\mathrm{E} g,\mathbf{k}}$ denoting the $g$th eigenmode.  The solution to the original PDEs (\ref{eq:psipde}) is therefore
\begin{equation}\label{eq:eigen2layer}
\boldsymbol{\psi}(\mathbf{x}, t) = \sum_{\mathbf{k}}^{\mathcal{K}}{ { e^{i\mathbf{k} \cdot \mathbf{x}}\hat{\boldsymbol{\psi}}_\mathbf{k}(t)}}  = \sum_{\mathbf{k}}^{\mathcal{K}}{ \sum_g^G{ e^{i\mathbf{k} \cdot \mathbf{x}}\hat{\psi}_{\mathrm{E} g,\mathbf{k}}(t)\mathbf{r}_{g,\mathbf{k}}}},
\end{equation}
where $\mathcal{K}$ is a finite set that contains the Fourier wavenumbers. $e^{i\mathbf{k}\cdot\mathbf{x}}$ is the two-dimensional Fourier basis function. The eigenvectors $\{\mathbf{r}_{g,\mathbf{k}}\}$ determine the weights of combining the eigenmodes.

With the eigenvectors in hand, we can use a linear stochastic model instead of the pure exponential $e^{\lambda_{g,\mathbf{k}}t}$ to better characterize the time evolution of $\hat {\psi}_{\mathrm{E} g,\mathbf{k}}$. The linear stochastic model that we adopt has the following form:
\begin{equation}\label{eq:ou}
\frac{\mathrm{d}\hat {\psi}_{\mathrm{E} g,\mathbf{k}} }{\mathrm{d}t}=  (-\gamma_{g,\mathbf{k}} + i\omega_{g,\mathbf{k}}) \hat \psi_{g,\mathbf{k}} + f_{g,\mathbf{k}} + \sigma_{g,\mathbf{k}} \dot{W}_{g,\mathbf{k}},
\end{equation}
where $-\gamma_{g,\mathbf{k}}$ and $\omega_{g,\mathbf{k}}$ represent the damping/dissipation and phase, respectively, which are associated with the eigenvalue. $f_{g,\mathbf{k}}$ is a constant forcing term. The stochastic noise $\sigma_{g,\mathbf{k}}\dot{W}_{g,\mathbf{k}}$ mimics the unresolved turbulent features originally caused by the nonlinearity. The model (\ref{eq:ou}) is known as the complex Ornstein–Uhlenbeck (OU) process \citep{gardiner2009stochastic}. The complex OU process can exactly match the mean and variance of data, and is a proper approximate model when the distribution of $\hat \psi_{\mathrm{E} g,\mathbf{k}}$ is nearly Gaussian. It has been applied as an effective surrogate forecast model in data assimilation \citep{farrell1993stochastic, berner2017stochastic, branicki2018accuracy, majda2018model, li2020predictability, harlim2008filtering, kang2012filtering}. %Finally, the deterministic solution is replaced by the stochastic one,

The CGDA is applied to $\hat{\psi}_{\mathrm{E} g,\mathbf{k}}(t)$ in the spectral space. The physical variables are reconstructed by the inverse Fourier transform. If we want to convert to the velocity, based on (\ref{eq:v_psi}) we have
\begin{equation}\label{eq:v_psik}
    \mathbf{v}_i (\mathbf{x}, t) =\sum_{\mathbf{k} }^{\mathcal{K}} e^{i\mathbf{k}\cdot\mathbf{x}}\hat{\psi}_{i,\mathbf{k}}(t)  \boldsymbol{\zeta}_{\mathbf{k}},
\end{equation}
where $\boldsymbol{\zeta}_{\mathbf{k}}=(ik_y,-ik_x)^\mathrm{T}$, and  $\hat{\psi}_{i,\mathbf{k}}$ is the Fourier coefficient of $i$th-layer stream function.

The motivation for using spectral decomposition and eigendecomposition is mainly to reduce computational costs. Spectral methods are commonly used by PDE solvers. Notably, it is natural to apply mode truncation with spectral methods and LSMs to reduce the computation burden since the time evolution of different modes is decoupled in the LSMs. Therefore, only a small subset of modes that contain significant energy is retained in the surrogate forecast model. Moreover, spectral representation avoids the expensive back-and-forth transformation between Eulerian and Lagrangian coordinates in Lagrangian data assimilation. By decoupling the original system through eigendecomposition, the coefficient matrix $\bm{\mathrm{a}}_1$ and noise level matrix $\bm{\mathrm{\Sigma}}_2$ becomes diagonal, which significantly accelerates the matrix manipulations in CGDA.

\section{Application to the two-layer quasi-geostrophic system}\label{sec:3}
In this section, we illustrate detailed ideas described in section~\ref{sec:da_multilayer} by giving an example application to the two-layer quasi-geostrophic (QG) equations. %Especially, we show that the highly nonlinear quasi-geostrophic turbulence can be effectively approximated by the linearized (\textcolor{red}{the 'linearized' may cause confusion}?)CGDA framework and demonstrate uniformly high skill in different computational regions. (\textcolor{red}{this is showed in the numerical results section?})

\subsection{Two-layer QG equations}
The two-layer quasi-geostrophic flow with topography reads \citep{qi2016low}
\begin{subequations}\label{eq:qg}
\begin{align}
\frac{\partial q_1}{\partial t} + J(\psi_1, q_1) + \beta \frac{\partial \psi_1}{\partial x} + U_1 \frac{\partial }{\partial x}\nabla^2 \psi_1 + \frac{k_d^2}{2} \left( U_1 \frac{\partial \psi_{2}}{\partial x} - U_{2} \frac{\partial \psi_1}{\partial x} \right) &=  - \nu \Delta^s q_1,    \\
\frac{\partial q_2}{\partial t} + J(\psi_2, q_2) + \beta \frac{\partial \psi_2}{\partial x} + U_2 \frac{\partial }{\partial x}\nabla^2 \psi_2 + \frac{k_d^2}{2} \left( U_2 \frac{\partial \psi_{1}}{\partial x} - U_{1} \frac{\partial \psi_2}{\partial x} \right) &= - \left( U_2 \frac{\partial h}{\partial x} + \kappa \nabla^2 \psi_2 \right) - \nu \Delta^s q_2.
\end{align}
\end{subequations}
The potential vorticity disturbance in the upper layer ($i = 1$) and lower layer ($i = 2$) is defined as
\begin{equation}\label{eq:q1q2}
q_1 = \nabla^2 \psi_1 + \frac{k_d^2}{2} (\psi_2 - \psi_1), \quad q_2 = \nabla^2 \psi_2 + \frac{k_d^2}{2} (\psi_1 - \psi_2) + h.
\end{equation}
Topography is introduced by $h$ in the lower layer equation. $(U_1, U_2)$ is the mean flow that can be decomposed as
\begin{equation}\label{eq:shearU}
U_1 = U_0 + U, \quad U_2 = U_0 - U,
\end{equation}
where $U_0$ is the constant mean structure and $U$ is the shear between the two layers. $J(A, B) = A_xB_y - A_yB_x$ represents the Jacobian operator. $k_d = \sqrt{8/{L_d}}$ is the baroclinic deformation wavenumber corresponding to the Rossby radius of deformation $L_d$. $\beta$ is the Rossby parameter. On the right-hand sides, the Ekman damping $\kappa \nabla^2 \psi_i$ only exists in the lower layer for the bottom friction, and hyperviscosity terms $ \nu \Delta^s q_i$ are added in both layers. For simplicity, we consider the case of equal layer depth, $H_1 = H_2 = H/{2}$, a large-scale topography $h$, and a zero mean flow $U_0=0$ and $U=const$. The boundary condition is periodic in both $x$ and $y$ directions.

\subsection{Reduced-order stochastic models}\label{sec:3b}
Here, we present details of developing reduced-order approximate models to the fully coupled two-layer QG model \eqref{eq:qg}. First, the LSMs of eigenmodes (as described in section \ref{sec:lsmda}), by keeping only linear terms of the flow equations, are proposed. These LSMs are used in the first step of the multi-step CGDA, which can also work as a standalone one-step method.  Since the tracers are directly driven by the surface layer, the observations play an important role in recovering the surface-layer flows, and therefore, a crude forecast model with linear approximations is often sufficient. Next, a conditional Gaussian approximation is proposed to better characterize the nonlinear coupling.

\subsubsection{LSMs of eigenmodes}\label{sec:ou}
To build LSMs of eigenmodes in spectral space for the two-layer QG equations,  we first linearize (\ref{eq:qg}) by neglecting the Jacobian terms. The linearized two-layer QG system has the following form:
\begin{subequations}\label{eq:linearqg}
 \begin{align}
\frac{\partial q_1}{\partial t}  + \beta \frac{\partial \psi_1}{\partial x} + U_1 \frac{\partial }{\partial x}\nabla^2 \psi_1 + \frac{k_d^2}{2} \left( U_1 \frac{\partial \psi_{2}}{\partial x} - U_{2} \frac{\partial \psi_1}{\partial x} \right) &=  - \nu \Delta^s q_1,    \\
\frac{\partial q_2}{\partial t}  + \beta \frac{\partial \psi_2}{\partial x} + U_2 \frac{\partial }{\partial x}\nabla^2 \psi_2 + \frac{k_d^2}{2} \left( U_2 \frac{\partial \psi_{1}}{\partial x} - U_{1} \frac{\partial \psi_2}{\partial x} \right) &= - \left( U_2 \frac{\partial h}{\partial x} + \kappa \nabla^2 \psi_2 \right) - \nu \Delta^s q_2,
\end{align}
\end{subequations}
 The viscous term $- \nu \Delta^s q_i$ and Ekman damping term $- \kappa \nabla^2 \psi_2$ are high-order terms and thus can be neglected. The constant forcing term $-  U_2 \partial h/{\partial x}$ is also neglected for the purpose of linear analysis. Plug (\ref{eq:q1q2}) into (\ref{eq:linearqg}), we obtain the equations with $\boldsymbol{\psi}$ being the only unknown
% {\footnotesize
\begin{subequations}\label{eq:qg_psi}
\begin{align}
&\frac{\partial}{\partial t} \left( \nabla^2 \psi_1 + \frac{k_d^2}{2} (\psi_2 - \psi_1) \right) + \beta \frac{\partial \psi_1}{\partial x} + U_1 \frac{\partial }{\partial x}\nabla^2 \psi_1 + \frac{k_d^2}{2} \left( U_1 \frac{\partial \psi_{2}}{\partial x} - U_{2} \frac{\partial \psi_1}{\partial x} \right) = 0,
\\
&\frac{\partial}{\partial t} \left( \nabla^2 \psi_2 + \frac{k_d^2}{2} (\psi_1 - \psi_2) \right) + \beta \frac{\partial \psi_2}{\partial x} + U_2 \frac{\partial }{\partial x}\nabla^2 \psi_2 + \frac{k_d^2}{2} \left( U_2 \frac{\partial \psi_{1}}{\partial x} - U_{1} \frac{\partial \psi_2}{\partial x} \right) = 0.
\end{align}
\end{subequations}
Applying the Fourier transform in the $\mathbf{x}$ variable leads to a system of ODEs for $\hat{\boldsymbol{\psi}}_{\mathbf{k}}$ in the form of (\ref{eq:psiode}) for each wavenumber $\mathbf{k}$. The eigenvalues and eigenvectors of the coefficient matrix $\hat{\bm{\mathsf{L}}}$ satisfy the following eigenequation
\begin{equation}\label{eq:eigen}
\begin{aligned}
     \lambda_\mathbf{k}\mathbf{r_k} &= -\bm{\mathsf{M}}_{\mathbf{k}}^{-1}\bm{\mathsf{N}}_{\mathbf{k}}\mathbf{r_k},  \\
    \bm{\mathsf{M}}_{\mathbf{k}}=\begin{bmatrix}
        -(|\mathbf{k}|^2+\frac{k_d^2}{2}),\frac{k_d^2}{2}\\
        \frac{k_d^2}{2}, -(|\mathbf{k}|^2+\frac{k_d^2}{2})
    \end{bmatrix},&\quad
    \bm{\mathsf{N}}_{\mathbf{k}}=k_x\begin{bmatrix}
       \beta -|\mathbf{k}|^2U+\frac{k_d^2}{2}U,\frac{k_d^2}{2}U\\
        -\frac{k_d^2}{2}U, \beta +|\mathbf{k}|^2U-\frac{k_d^2}{2}U
    \end{bmatrix}.
\end{aligned}
\end{equation}
Solving this $2\times 2$ eigenequations gives two eigenvalues and two eigenvectors
\begin{equation}\label{eq:eigensolution}
\begin{aligned}
   \lambda_{\{1,2\},\mathbf{k}}=& \frac{k_x \left(\beta (k_d^2 + 2 |\mathbf{k}|^2) \pm  \sqrt{\beta^2 k_d^4 + 4 |\mathbf{k}|^4 (-k_d^4 + |\mathbf{k}|^4) U^2}\right)}{2 |\mathbf{k}|^2 (k_d^2 + |\mathbf{k}|^2)}, \\
%\omega_{2,\mathbf{k}}=&\frac{k_x \left(\beta (k_d^2 + 2 |\mathbf{k}|^2) - \sqrt{\beta^2 k_d^4 + 4 |\mathbf{k}|^4 (-k_d^4 + |\mathbf{k}|^4) U^2}\right)}{2 |\mathbf{k}|^2 (k_d^2 + |\mathbf{k}|^2)},     \\
\mathbf{r}_{\{1,2\},\mathbf{k}} &=\left(-\frac{2U|\mathbf{k}|^4 \mp \sqrt{\beta^2 k_d^4 + 4 |\mathbf{k}|^4 (-k_d^4 + |\mathbf{k}|^4) U^2}}{k_d^2 (\beta - 2 |\mathbf{k}|^2 U)},1\right)^{\mathrm{T}}, \\
%\mathbf{r}_{2,\mathbf{k}} &=\left(-\frac{2U|\mathbf{k}|^4  + \sqrt{\beta^2 k_d^4 + 4 |\mathbf{k}|^4 (-k_d^4 + |\mathbf{k}|^4) U^2}}{k_d^2 (\beta - 2 |\mathbf{k}|^2 U)}, 1\right)^{\mathrm{T}}.
\end{aligned}
\end{equation}
where the subscripts $1,2$ is corresponding to the positive or negative sign before the square root.

We can continue to develop LSMs and obtain the stochastic solution as described in Section \ref{sec:lsmda}. The resulting LSMs can be written in the following form
\begin{equation}\label{eq:lsm_eg}
    \frac{\mathrm{d}\boldsymbol{\Psi}_\mathrm{E}}{\mathrm{d}t} = (-\bm{\mathsf{\Gamma}_\mathrm{E}} + i\bm{\mathsf{\Omega}}_\mathrm{E})\boldsymbol{\Psi}_\mathrm{E} + \mathbf{F}_\mathrm{E}+ \bm{\mathsf{\Sigma}}_\mathrm{E} \dot{\mathbf{W}}_\mathrm{E},
\end{equation}
where $\boldsymbol{\Psi}_\mathrm{E} =[\cdots \hat{\psi}_{\mathrm{E}1,\mathbf{k}}, \cdots \hat{\psi}_{\mathrm{E}2,\mathbf{k}},\cdots]^{\mathrm{T}}$ is a $2|\mathcal{K}| \times 1$ column vector that contains the Fourier coefficients of the two eigenmodes, with $|\mathcal{K}|$ denoting the number of elements in set $\mathcal{K}$. $\bm{\mathsf{\Gamma}}_\mathrm{E}$ and $\bm{\mathsf{\Omega}}_\mathrm{E}$ are diagonal matrices corresponding to the damping and phase coefficients, respectively. $\mathbf{F}_\mathrm{E}$ is the forcing vector. $\bm{\mathsf{\Sigma}}_\mathrm{E}$ is a diagonal matrix of noise coefficients.

\subsubsection{Conditional Gaussian nonlinear stochastic model}\label{sec:3b2}
Next, we develop a conditional Gaussian nonlinear model serving as the forecast model in the second data assimilation step for the two-layer QG system. Instead of completely neglecting the nonlinear coupling term as in \eqref{eq:linearqg}, we only drop the quadratic terms of lower layer $\psi_2$ in (\ref{eq:qg}) while keeping the remaining nonlinear terms. This leads to the following nonlinear model with the conditional Gaussian structure
\begin{subequations}\label{eq:cg_raw}
\begin{align}
\frac{\partial q_1}{\partial t} + J(\psi_1, q_1) + \beta \frac{\partial \psi_1}{\partial x} + U_1 \frac{\partial }{\partial x}\nabla^2 \psi_1 + \frac{k_d^2}{2} \left( U_1 \frac{\partial \psi_{2}}{\partial x} - U_{2} \frac{\partial \psi_1}{\partial x} \right) &= - \nu \Delta^s q_1, \\
\frac{\partial {q}_2}{\partial t} + J(\psi_2, \frac{k_d^2}{2} \psi_1 + h) + \beta \frac{\partial \psi_2}{\partial x} + U_2 \frac{\partial }{\partial x}\nabla^2 \psi_2 + \frac{k_d^2}{2} \left( U_2 \frac{\partial \psi_{1}}{\partial x} - U_{1} \frac{\partial \psi_2}{\partial x} \right) &= -  U_2 \frac{\partial h}{\partial x} - \kappa \nabla^2 \psi_2 - \nu \Delta^s q_2.
\end{align}
\end{subequations}
The modified system (\ref{eq:cg_raw}) fits into the CGNS framework \eqref{eq:cgns}. Applying discrete Fourier transform to $\psi_1$ and $\psi_2$, we have
\begin{subequations}\label{eq:cg_qg_spec}
    \begin{align}
        \frac{\partial q_{1,\mathbf{k}}}{\partial t}  &=  -ik_x\left((\beta -U_1|\mathbf{k}|^2-\frac{k_d^2}{2}U_2)\psi_{1,\mathbf{k}} + \frac{k_d^2}{2}U_1\psi_{2,\mathbf{k}}\right) - J_{\mathbf{k}}(\psi_1, q_1) -\nu |\mathbf{k}|^{2s} q_{1,\mathbf{k}},
        \\ \frac{\partial q_{2,\mathbf{k}}}{\partial t} &= -ik_x\left((\beta -U_2|\mathbf{k}|^2-\frac{k_d^2}{2}U_1)\psi_{2,\mathbf{k}} + \frac{k_d^2}{2}U_2\psi_{1,\mathbf{k}} + U_2h_{\mathbf{k}}\right) + \kappa|\mathbf{k}|^2\psi_{2,\mathbf{k}} - J_{\mathbf{k}}(\psi_2, \tilde{q}_2) -\nu |\mathbf{k}|^{2s} q_{2,\mathbf{k}},
    \end{align}
\end{subequations}
where $\psi_{i,\mathbf{k}}=\hat{\psi}_{i,\mathbf{k}}(t)e^{i\mathbf{k}\cdot \mathbf{x}}$ is the Fourier mode of $\psi_i$. The Fourier modes of $q_1$ and $q_2$ are
\begin{equation}
    \begin{aligned}
        q_{1,\mathbf{k}} &= -(|\mathbf{k}|^2+\frac{k_d^2}{2})\psi_{1,\mathbf{k}} + \frac{k_d^2}{2}\psi_{2,\mathbf{k}}, \\
        q_{2,\mathbf{k}} &= - (|\mathbf{k}|^2+\frac{k_d^2}{2})\psi_{2,\mathbf{k}} + \frac{k_d^2}{2}\psi_{1,\mathbf{k}}  + h_{\mathbf{k}}.
    \end{aligned}\label{eq:converion}
\end{equation}
The Jacobian term $J_{\mathbf{k}}(\psi, q)$ is given by assigning summands in $J(\psi,q)$ to the $\mathbf{k}$\textit{th} mode
\begin{equation}
    J_{\mathbf{k}}(\psi, q)=\sum^{\mathcal{K},\mathcal{K}}_{\substack{\mathbf{m},\mathbf{n} \\ \mathbf{m} +\mathbf{n}=\mathbf{k}}} (m_xn_y-m_yn_x)
        \psi_{\mathbf{n}}q_{\mathbf{m}}.\label{eq:jacobian}
\end{equation}
(\ref{eq:cg_qg_spec}) can be written in the form of
\begin{equation}\label{eq:cg_qg_mat}
\begin{aligned}
    \begin{bmatrix}
        -(|\mathbf{k}|^2+\frac{k_d^2}{2}),\frac{k_d^2}{2}\\
        \frac{k_d^2}{2}, -(|\mathbf{k}|^2+\frac{k_d^2}{2})
    \end{bmatrix}
    \begin{bmatrix}
        \frac{\partial}{\partial t}\psi_{1,\mathbf{k}} \\ \frac{\partial}{\partial t}\psi_{2,\mathbf{k}}
    \end{bmatrix}
    = \begin{bmatrix}
        \mathrm{RHS}_1 \\ \mathrm{RHS}_2
    \end{bmatrix}.
\end{aligned}
\end{equation}
Diagonalizing the coefficient matrix with Gaussian elimination, and plugging in $\mathrm{RHS}_1, \mathrm{RHS}_2$, and (\ref{eq:shearU}), we obtain
\begin{subequations}\label{eq:cg_qg}
\small{\begin{align}
\frac{\partial \psi_{1,\mathbf{k}}}{\partial t} =& C_{\mathbf{k}}\left(ik_x \left(\left((|\mathbf{k}|^2+\frac{k_d^2}{2})\beta - |\mathbf{k}|^4U \right) \psi_{1,\mathbf{k}} +
        (\frac{k_d^2}{2} \beta + k_d^2|\mathbf{k}|^2U)\psi_{2,\mathbf{k}} - \frac{k_d^2}{2}Uh_{\mathbf{k}} \right)
        + \tilde{D}_{1,\mathbf{k}}
        + \tilde{J}_{1,\mathbf{k}} \right),\\
\frac{\partial \psi_{2,\mathbf{k}}}{\partial t} =&
        C_{\mathbf{k}}\left(ik_x \left((\frac{k_d^2}{2}\beta - k_d^2|\mathbf{k}|^2U )\psi_{1,\mathbf{k}} +
        \left((|\mathbf{k}|^2+\frac{k_d^2}{2}) \beta + |\mathbf{k}|^4U \right)\psi_{2,\mathbf{k}} - (|\mathbf{k}|^2+\frac{k_d^2}{2})Uh_{\mathbf{k}} \right)
        + \tilde{D}_{2,\mathbf{k}}
        + \tilde{J}_{2,\mathbf{k}} \right),
    \end{align}}
\end{subequations}
where $C_{\mathbf{k}}=1/\left({|\mathbf{k}|^2(|\mathbf{k}|^2+k_d^2)}\right)$ is a normalization constant. $\tilde{D}_{i,\mathbf{k}} $ and $\tilde{J}_{i,\mathbf{k}}$ correspond to the damping and Jacobian terms, with exact forms given in appendix C. 

%We are just one step away from fitting the modified two-layer QG into the conditional Gaussian framework. 
The final step is to plug $\psi_{\mathbf{k}}=\hat{\psi}_{\mathbf{k}}(t)e^{i\mathbf{k}\cdot \mathbf{x}}$ into (\ref{eq:cg_qg}) and add stochastic terms. Through basic algebraic manipulations, we will arrive where $\partial \cdot /{\partial t}$ becomes $\mathrm{d} \cdot/{\mathrm{d} t}$. This gives rise to the conditional Gaussian stochastic model for the QG:
\begin{subequations}\label{eq:cg_2layer}
    \begin{align}
\frac{\mathrm{d \boldsymbol{\Psi}_1}}{\mathrm{d} t} &= \mathbf{A}_0(\boldsymbol{\Psi}_1,t) + \bm{\mathsf{A}}_1(\boldsymbol{\Psi}_1,t) \boldsymbol{\Psi}_2  + \bm{\mathsf{\Sigma}}_1 \dot{\mathbf{W}}_1, \\
\frac{\mathrm{d}\boldsymbol{\Psi}_2}{\mathrm{d}t} &= \mathbf{a}_0(\boldsymbol{\Psi}_1,t) + \bm{\mathsf{a}}_1(\boldsymbol{\Psi}_1,t) \boldsymbol{\Psi}_2  + \bm{\mathsf{\Sigma}}_2 \dot{\mathbf{W}}_2,
    \end{align}
\end{subequations}
where $\boldsymbol{\Psi}_{i} =[\cdots \hat{\psi}_{i,\mathbf{k}}, \cdots ]^{\mathrm{T}}$ is a $|\mathcal{K}| \times 1$ column vector that contains the $i$th layer's Fourier coefficients. $\mathbf{A}_0(\boldsymbol{\Psi}_1,t)$ and $\mathbf{a}_0(\boldsymbol{\Psi}_1,t)$ are $|\mathcal{K}| \times 1$ column vectors. $\bm{\mathsf{A}}_1(\boldsymbol{\Psi}_1,t)$ and $\bm{\mathsf{a}}_1(\boldsymbol{\Psi}_1,t)$ are $|\mathcal{K}|\times |\mathcal{K}|$ matrices. The exact forms of $\mathbf{A}_0$, $\mathbf{a}_0$, $\bm{\mathsf{A}}_1$, and $\bm{\mathsf{a}}_1$ are given in appendix C. $\bm{\mathsf{\Sigma}}_1 \dot{\mathbf{W}}_1$ and $\bm{\mathsf{\Sigma}}_1 \dot{\mathbf{W}}_2$ are stochastic noises that compensate the unresolved terms.

\subsubsection{Model calibration and evaluation}
For calibrating the LSMs (\ref{eq:lsm_eg}), four parameters --- damping $\bm{\mathsf{\Gamma}}_\mathrm{E}$, phase $\bm{\mathsf{\Omega}}_\mathrm{E}$, forcing $\mathbf{F}_\mathrm{E}$, and noise strength $\bm{\mathsf{\Sigma}}_\mathrm{E}$, must be calibrated by matching four statistics of data --- mean, variance, real and imaginary parts of the decorrelation time \citep{chen2023stochastic}. For calibrating the conditional Gaussian stochastic models (\ref{eq:cg_2layer}), we recompute all the coefficients using linear regression, which is equivalent to adding closures. In this paper, we assume the deterministic coefficients are known, only calibrating the stochastic noise strength $\bm{\mathsf{\Sigma}}_1$ and $\bm{\mathsf{\Sigma}}_2$ by matching one-step residual variances. See appendices A\ref{Appendix:Aa} and A\ref{Appendix:Ab} for more details about the model calibration.

We use several statistical metrics to evaluate the calibration, specifically for the QG system. They are the kinetic energy (KE), the available potential energy (APE), the total energy (E), and the enstrophy ($\mathcal{E}$). See more details in appendix A\ref{Appendix:Ac} about how these metrics are defined and calculated. Notably, the LSMs of eigenmodes with a diagonal $\bm{\mathsf{\Sigma}}_{\mathrm{E}}$ may fail to capture the energy of layer flows, even though they are well-calibrated. Because a diagonal $\bm{\mathsf{\Sigma}}_{\mathrm{E}}$ means the noises of different eigenmodes are independent of each other. However, the correlations among noises can exist in realistic data from nonlinear QG equations (see appendix A\ref{Appendix:Ad} for the conversion between eigenmodes energy and layers energy). This can be solved by allowing correlated noises in the LSMs, which improves the linear modeling. Note that in such a case, $\bm{\mathsf{\Sigma}}_{\mathrm{E}}$ is no longer diagonal, which increases computational burden in data assimilation. More details of the LSMs with correlated noises are put in Appendix B.

\subsection{Lagrangian CGDA for the two-layer QG model}
Finally, we construct the CGNSs and give the analytic formulae for solving posterior statistics, which are used in the one-step and the multi-step CGDA. %combining the passive tracer observation and the approximated flow system using the one-step and multi-step approaches.
\subsubsection{One-step CGDA with linear stochastic flow Model}
The Lagrangian tracer equation (\ref{eq:multilayera}) together with the linear stochastic flow model (\ref{eq:lsm_eg}) form a conditional Gaussian system:
\begin{subequations}
\begin{align}
\text{Observations:}\ \  \frac{\mathrm{d}\mathbf{X}}{\mathrm{d}t} &= \bm{\mathsf{A}}(\mathbf{X},t) \boldsymbol{\Psi}_\mathrm{E}+ \bm{\mathsf{\Sigma}}_\mathbf{X} \dot{\mathbf{W}}_\mathbf{X}, \label{eq:tracer}\\
\text{Underlying flow:} \ \ \frac{\mathrm{d}\boldsymbol{\Psi}_\mathrm{E}}{\mathrm{d}t} &= \bm{\mathsf{a}}\boldsymbol{\Psi}_\mathrm{E} + \mathbf{F}_\mathrm{E} + \bm{\mathsf{\Sigma}}_\mathrm{E} \dot{\mathbf{W}}_\mathrm{E} ,\label{eq:flow_eigen}
\end{align}
\end{subequations}
where $\mathbf{X} = (\ldots, x_l,\ldots, y_l,\ldots)^{\mathrm{T}}$ is a \(2L \times 1\) dimensional vector containing the displacements of all the Lagrangian tracers. The matrix $\bm{\mathsf{a}}=-\bm{\mathsf{\Gamma}}_\mathrm{E}  + i\bm{\mathsf{\Omega}}_\mathrm{E} $. Assuming observation noises are independent, $\bm{\mathsf{\Sigma}}_\mathbf{X}$  is a diagonal matrix. Particularly, $\bm{\mathsf{A}}(\mathbf{X},t)$ is a $2L\times 2|\mathcal{K}|$ matrix given by
\begin{equation}\label{eq:eigen2v}
\begin{aligned}
    \bm{\mathsf{A}} &= \bm{\mathsf{Z}}\bm{\mathsf{H}} \\
    with \ \bm{\mathsf{Z}}=\begin{bmatrix} 
     & \vdots &  \\
        \cdots & e^{i\mathbf{k}\cdot \mathbf{x}_l} {\zeta}_{\mathbf{k}}^{(1)} & \cdots\\
        & \vdots & \\
        \cdots & e^{i\mathbf{k}\cdot \mathbf{x}_l} {\zeta}_{\mathbf{k}}^{(2)} & \cdots\\
      & \vdots & 
    \end{bmatrix}, 
    \bm{\mathsf{H}} &=\begin{bmatrix}
        \ddots &  &   & \ddots &  &  \\
         & r_{1,\mathbf{k}}^{(1)} &  & &r_{2,\mathbf{k}}^{(1)} &  \\
          &  & \ddots &  & & \ddots 
    \end{bmatrix}
\end{aligned}
\end{equation}
where $\bm{\mathsf{Z}}$ is a $2L\times |\mathcal{K}|$ matrix with ($\bm{\mathsf{Z}}_{l,\mathbf{k}},\bm{\mathsf{Z}}_{l+L,\mathbf{k}})^\mathrm{T}=e^{i\mathbf{k}\cdot \mathbf{x}_l} \boldsymbol{\zeta}_{\mathbf{k}}$. $\bm{\mathsf{H}}$ is a $|\mathcal{K}|\times2|\mathcal{K}|$ matrix with $\bm{\mathsf{H}}_{\mathbf{k},\mathbf{k}}=r_{1,\mathbf{k}}^{(1)}, \bm{\mathsf{H}}_{\mathbf{k},\mathbf{k}+|\mathcal{K}|} = r_{2,\mathbf{k}}^{(1)}$ and other entries being 0. The superscript $^{(1)}$ and $^{(2)}$ denote the first and second elements of a vector, respectively. Note that $\bm{\mathsf{H}}$ transform the eigenmodes to the first layer, as $\bm{\mathsf{H}}\boldsymbol{\Psi}_\mathrm{E}=\boldsymbol{\Psi}_1$, where $\boldsymbol{\Psi}_1=[\cdots \hat{\psi}_{1,\mathbf{k}},\cdots]^{\mathrm{T}}$ contains the Fourier coefficients of the first-layer stream function, and $\bm{\mathsf{Z}}$ does the conversion from stream function to velocity and inverse Fourier transform. This can be derived from (\ref{eq:v_psik}).

The analytic formulae for solving the posterior mean and covariance  of $\boldsymbol{\Psi}_\mathrm{E}(t) $ given tracer trajectory observation $\mathbf{X}(s \leq t) $ are as follows
\begin{subequations}\label{eq:lsmeg_da}
\begin{align}
    \frac{\mathrm{d}\Bar{\boldsymbol{\Psi}}_\mathrm{E}}{\mathrm{d}t}  =  (\mathbf{F}_\mathrm{E} + \bm{\mathsf{a}} \Bar{\boldsymbol{\Psi}}_\mathrm{E}) + (\bm{\mathsf{R}}\bm{\mathsf{A}}^* )(\bm{\mathsf{\Sigma}}_\mathbf{X} \bm{\mathsf{\Sigma}}_\mathbf{X} ^*)^{-1}
        \left(\frac{\mathrm{d}\mathbf{X}}{\mathrm{d}t}-\bm{\mathsf{A}}\Bar{\boldsymbol{\Psi}}_\mathrm{E}\right), \label{eq:mean}\\
      \frac{\mathrm{d}\bm{\mathsf{R}}_\mathrm{E}}{\mathrm{d}t}  = \bm{\mathsf{a}}\bm{\mathsf{R}}_\mathrm{E}+\bm{\mathsf{R}}_\mathrm{E}\bm{\mathsf{a}}^*+{\bm{\mathsf{\Sigma}}_{\boldsymbol{\Psi}_\mathrm{E}}}{\bm{\mathsf{\Sigma}}}_{\boldsymbol{\Psi}_\mathrm{E}}^*
    - (\bm{\mathsf{R}}_\mathrm{E}\bm{\mathsf{A}}^*) (\bm{\mathsf{\Sigma}}_\mathbf{X} \bm{\mathsf{\Sigma}}_\mathbf{X} ^*)^{-1} (\bm{\mathsf{A}}{\bm{\mathsf{R}}_\mathrm{E}}^*).
\end{align}
\end{subequations}

\subsubsection{Multi-step CGDA}
Recall that the CGDA based on linear stochastic flow models serves as the first step of multi-step CGDA. After getting the posterior of surface-layer flow via (\ref{eq:lsmeg_da}), we can sample pseudo-observations from the posterior. Here we use a backward sampling formula \citep{chen_efficient_2020}
\begin{equation}\label{eq:backsample}
    \frac{\overset{\leftarrow}{\mathrm{d} \boldsymbol{\Psi}_\mathrm{E}}}{\mathrm{d}t}  = \left(-\mathbf{F}_\mathrm{E} -\bm{\mathsf{a}} \boldsymbol{\Psi}_\mathrm{E}\right) + \bm{\mathsf{\Sigma}}_{\mathrm{E}}\bm{\mathsf{\Sigma}}_{\mathrm{E}}^*\bm{\mathsf{R}}_\mathrm{E}^{-1}(\Bar{\boldsymbol{\Psi}}_\mathrm{E}- \boldsymbol{\Psi}_\mathrm{E})+ \bm{\mathsf{\Sigma}}_{\mathrm{E}}\dot{\mathbf{W}}_{\mathrm{E}},
\end{equation}
where $\overset{\leftarrow}{\mathrm{d} \cdot}/\mathrm{d}t$ denotes the backward derivative, i.e., the negative of the usual derivative. The formula (\ref{eq:backsample}) runs backward from $t=T$ to $t=0$, with initial samples drawn from the posterior $\mathcal{N}\left(\Bar{\boldsymbol{\Psi}}_\mathrm{E}(T), \bm{\mathsf{R}}_\mathrm{E}(T)\right)$. It is the optimal backward sampling formula for the unobserved variables in CGDA \citep{chen_efficient_2020}. The conversion form eigenmodes to upper layer is realized through $\bm{\mathsf{H}}\boldsymbol{\Psi}_\mathrm{E}=\boldsymbol{\Psi}_1$ , where $\bm{\mathsf{H}}$ is defined in (\ref{eq:eigen2v}).

With pseudo-observations of the upper layer,  we perform CGDA based on the nonlinear conditional Gaussian stochastic model (\ref{eq:cg_2layer}) for each pseudo-observation $\boldsymbol{\Psi}_1^{(n)}$ as described in (\ref{eq:cgposterior}). The posterior mean and covariance of the lower layer for the $n$th sample are solved by the analytic formulae (\ref{eq:cg_meanvar}), with $\mathbf{u}_1$ and $\mathbf{u}_2$ replaced by  $\boldsymbol{\Psi}_1^{(n)}$ and $\boldsymbol{\Psi}_2^{(n)}$. The ultimate lower-layer posterior combining all samples is a mixture of Gaussians according to (\ref{eq:montecarlo}).

\subsection{Experimental design}
The multi-step CGDA and the one-step CGDA with LSMs are tested in the two-layer QG problem. The QG model runs using the fourth order Runge Kutta pseudo-spectral scheme for $N_t=205,000$ time steps with a time step $\mathrm{d}t=0.002$, in a double periodic domain $[-\pi, \pi)^2$ with $K=128$ grid points (also Fourier modes) in each direction, so the total number of grid points is $K^2=16384$. The initial $5000$ time steps are excluded from the analysis to allow for spin-up. The QG simulations serve as the ground truth, with default parameters outlined in Table \ref{qgparams}. Synthetic tracer observations are generated based on the true ocean field at every time step, with independent observation noises of strength $\sigma_x=\sigma_y=0.1$. Unless specified otherwise, $L=256$ tracer observations are used in data assimilation. The data assimilation experiments start from an equilibrium state and run for 200,000 time steps. There is no initial error or uncertainty. The reduced-order stochastic models are calibrated on QG simulations. Only the Fourier modes within a truncation radius $r=16$ from the (0,0) mode (797 modes in total) are integrated and filtered in data assimilation. By default, the multi-step CGDA uses $N_{s}=16$ ensembles.
\begin{table}[h]
\caption{Default non-dimensional parameters of the two-layer QG model.}\label{qgparams}
\begin{center}
\begin{tabular}{ccccrrcr}
\topline
 Deformation wavenumber & $k_d=10$ \\
 Rossby parameter  & $\beta=22$  \\
 Zonal mean flow & $U_0=0$  \\
 Zonal shear flow & $U=1$  \\
 Ekman damping & $\kappa=9$ \\
 Hyperviscosity & $\nu=10^{-12}$\\
 Topography & $h=40\big(\cos(x)+2cos(2y)\big)$ \\
\botline
\end{tabular}
\end{center}
\end{table}

% \begin{table}[h]
% \caption{Default experiment parameters.}\label{qgparams}
% \begin{center}
% \begin{tabular}{ccccrrcr}
% \topline
%  Number of time steps & $N_t=205000$ \\
%  Time step & $dt=0.002$ \\
%  Domain & $[-\pi, \pi)^2$ \\
%  Grid points (Fourier modes) & $K_x=K_y=128$ \\
%  Spectral truncation radius & $r=16$ \\
%  Observation noise strength & $\sigma_x=\sigma_y=0.1$ \\
%  Number of tracers & $L=256$ \\
%  Ensemble size & $N_s=16$ \\
%  Deformation wavenumber & $k_d=10$ \\
%  Rossby parameter  & $\beta=22$  \\
%  Zonal mean flow & $U_0=0$  \\
%  Zonal shear flow & $U=1$  \\
%  Ekman damping & $\kappa=9$ \\
%  Hyperviscosity & $\nu=10^{-12}$\\
%  Topography & $h=40\big(cos(x)+2cos(2y)\big)$ \\
% \botline
% \end{tabular}
% \end{center}
% \end{table}

The root-mean-square error (RMSE) is the metric used for evaluation, which is defined as
\begin{equation*}
    \text{RMSE} = \sqrt{\frac{1}{K^2} \sum_{k_y=1}^{K}\sum_{k_x=1}^{K} (\psi^{a}_{(k_x, k_y)} - \psi^{t}_{(k_x, k_y)})^2},
\end{equation*}
where the subscripts $k_x$ and $k_y$ denote the grid index in $x$ and $y$ direction, respectively. The superscript $a$ denotes analysis, the quantity to be evaluated. And the superscript $t$ denotes the truth. For time series, the time mean of RMSEs is used as the single scalar metric for evaluation.

\begin{figure}[h]
 \centerline{\includegraphics[width=27pc]{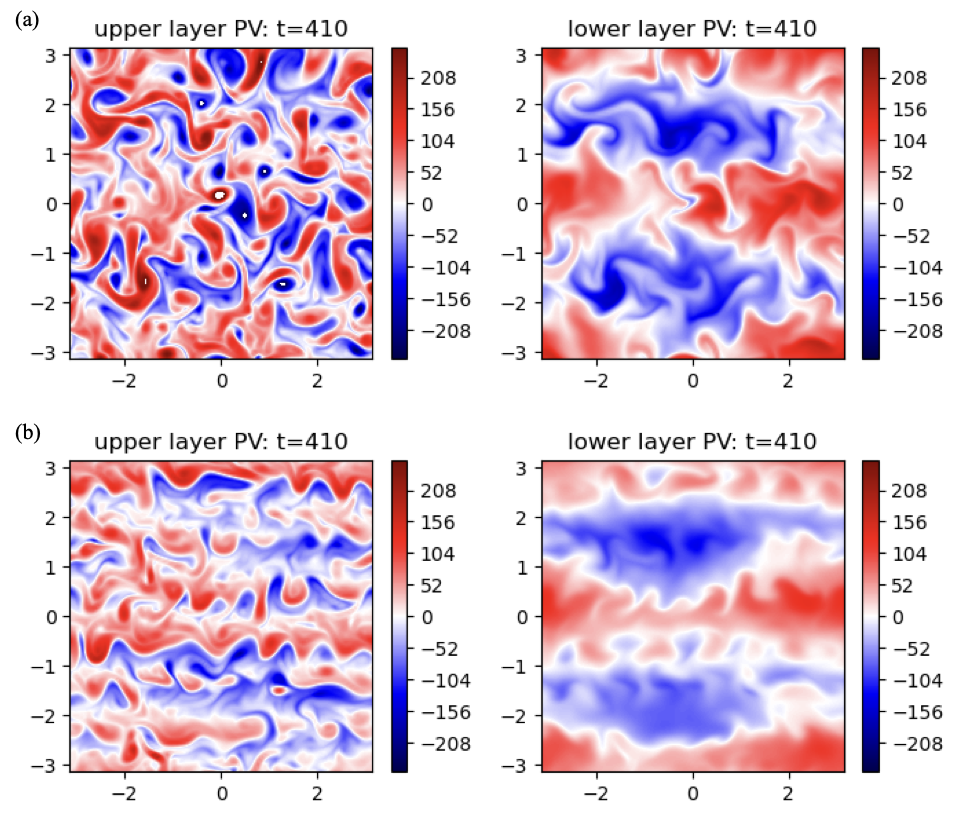}}
  \caption{Snapshots of the potential vorticity field of QG simulations for (a) strongly turbulent regime $\beta=22$ and (b) moderately turbulent regime $\beta=111$.}\label{fig:regimes}
\end{figure}

The data assimilation methods are evaluated in two regimes of the QG system. As $\beta$ decreases, the QG system becomes more turbulent (Fig.~\ref{fig:regimes}). We choose $\beta=111$ and $\beta=22$, which result in moderately turbulent and strongly turbulent flow structures, respectively. Snapshots of the typical vorticity field of the two test cases are shown in Fig.~\ref{fig:regimes}, illustrating complex flows with distinctive features. Notably, the linearized QG system (\ref{eq:linearqg}) becomes unstable when $\beta$ gets larger. In contrast, the modified conditional-Gaussian-QG system (\ref{eq:cg_raw}) is stable for all choices of $\beta$, because the dropped quadratic terms in the conditional-Gaussian-QG system do not break the total energy conservation in the original QG system \citep{vallis_geostrophic_2017}.

Sensitivities of data assimilation methods to the ensemble size $N_s$ and tracers number $L$ are tested using the default parameters configuration. In addition to the data assimilation experiments with surface tracer observations, we also compare the nonlinear conditional Gaussian flow model to the linear flow model in both free forecasts and data assimilation with direct upper-layer observations as a complementary demonstration of the benefits of incorporating additional quadratic terms.  Finally, we test the multi-step CGDA with a constant posterior covariance, which is estimated from long-term statistics, similar to the three-dimensional variational method \citep[3DVAR;][]{lorenc_analysis_1986}. Using constant covariance can dramatically reduce computational costs. The detailed calibration results are described in appendixes A and B.

\section{Numerical Results}

\subsection{Comparison of the one-step CGDA and the multi-step CGDA in the strongly turbulent regime}
\begin{figure}[h]
 \centerline{\includegraphics[width=33pc]{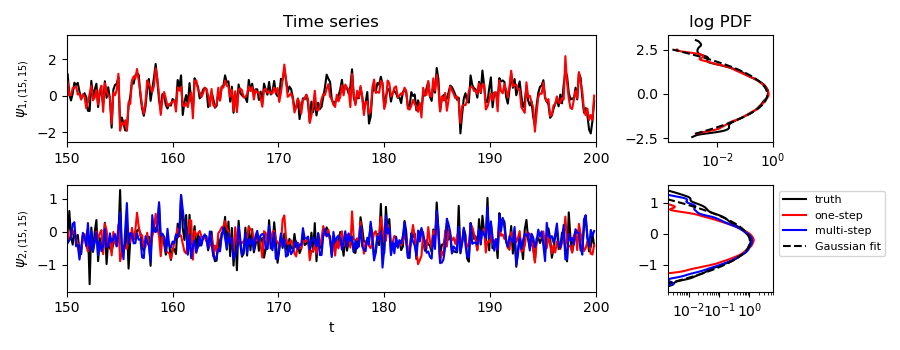}}
  \caption{Trajectories and the corresponding log-scale pdfs of $\psi_1$ and $\psi_2$ at displacement $(x,y)=(-2.4,-2.4)$. Black lines are true signals from QG. Red lines are one-step CGDA posterior. Blue lines are multi-step CGDA posterior. Black dash lines are fitted Gaussian pdfs of the truth. Only a window of time $150<t<200$ is shown for the trajectories.}\label{fig:layer_beta22}
\end{figure}

\begin{figure}[h]
 \centerline{\includegraphics[width=27pc]{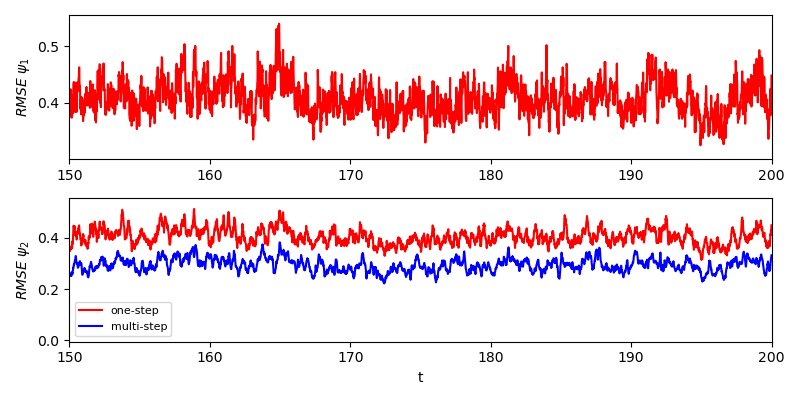}}
  \caption{Posterior RMSE series of one-step CGDA (red) and multi-step CGDA (blue) for $\psi_1$ and $\psi_2$. Only a window of time $150<t<200$ is shown.}\label{fig:rmses_beta22}
\end{figure}

For the default configuration, Fig. \ref{fig:layer_beta22} shows the trajectories and log-scale PDFs of the upper- and lower-layer stream functions at one physical location. The one-step CGDA using a linear flow model well recovers the upper-layer flow with 256 tracers. Compared to the one-step CGDA, the multi-step CGDA that incorporates a nonlinear flow model better recovers the lower-layer flow, as reflected both in the trajectories and PDFs. This is further confirmed by the RMSE in the posterior time series (Fig. \ref{fig:rmses_beta22}). The multi-step CGDA has a posterior mean RMSE of 0.291 for $\psi_2$, significantly smaller than the one-step CGDA's mean RMSE of 0.400.

\begin{figure}[htbp!]
 \centerline{\includegraphics[width=27pc]{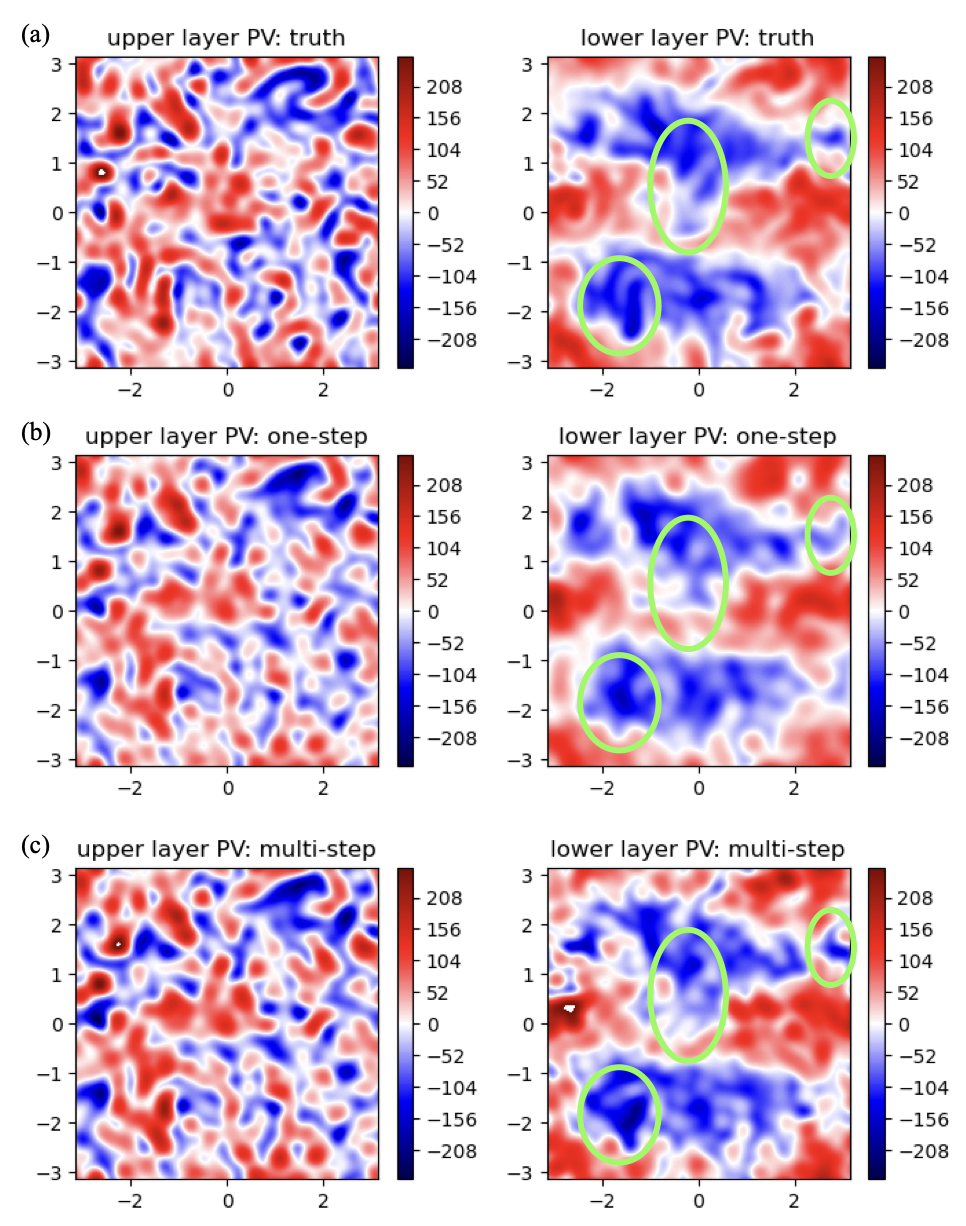}}
  \caption{Snapshots of the potential vorticity field constructed from truncated Fourier modes with truncation radius $r=16$. (a) truth, (b) one-step CGDA posterior, and (c) multi-step CGDA posterior. The left column is the upper layer and the right column is the lower layer.}\label{fig:flowsnap}
\end{figure}

\begin{figure}[htbp!]
 \centerline{\includegraphics[width=39pc]{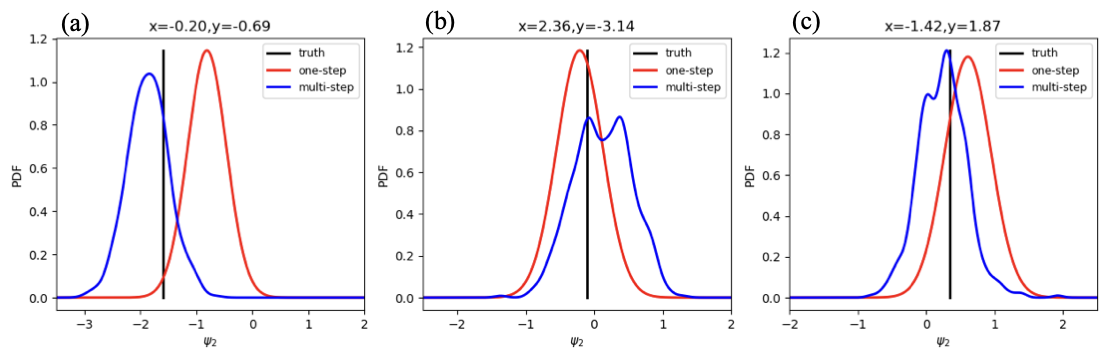}}
  \caption{Snapshots of the posterior pdfs at three different displacements, (a) nearly Gaussian case, (b) bimodal case, and (c) skewed case. The truth is marked with a black straight line. The one-step CGDA posterior (red) is Gaussian. The multi-step CGDA posterior (blue) is a mixture of Gaussians, and is smoothed with Gaussian kernel density estimation.}\label{fig:mog}
\end{figure}

Figure \ref{fig:flowsnap} displays the snapshots of the potential velocity field. The truth is constructed from the truncated Fourier modes as the data assimilation posterior for comparison. The multi-step CGDA better recovers the flow, revealing finer structures and small-scale features compared to the one-step CGDA, as highlighted by the green circles. We further inspected the posterior distributions of different data assimilation methods. Figure \ref{fig:mog} shows the snapshots of posterior PDFs at three different displacements. For case (a), both the one-step CGDA and multi-step CGDA have a (nearly) Gaussian posterior. In contrast, for cases (b) and (c), the multi-step CGDA gives either a bimodal or skewed posterior, demonstrating its capability to capture non-Gaussian posterior distributions with a mixture of Gaussians.

\subsection{Comparison of data assimilation skills in the two test regimes}
Consistent results are obtained in the moderately turbulent regime $\beta=111$. Table \ref{tab:rmse_tracer} summarizes the posterior mean RMSEs of data assimilation methods in the two regimes. The multi-step CGDA has smaller posterior mean RMSEs than the one-step CGDA in both regimes. Notably, both data assimilation methods achieve smaller posterior errors in the moderately turbulent regime $\beta=111$ than in the strongly turbulent regime $\beta=22$. The performance difference between the one-step CGDA and the multi-step CGDA shrinks as the flow becomes less turbulent, where linear structures become more dominant, as expected. Therefore, the advantages of multi-step CGDA over one-step CGDA are more pronounced in more intermittently unstable and turbulent flows.
% \begin{table}[h]
% \caption{Time mean RMSEs of one-step CGDA and multi-step CGDA with different $\beta$.}\label{tab:rmse_tracer}
% \begin{center}
% \begin{tabular}{ccc}
% \topline
% Regime & $\beta=22$ & $\beta=111$ \\
% \midline
% One-step CGDA & 0.400 & 0.137  \\
% Multi-step CGDA & 0.291 & 0.113 \\
% Multi-step CGDA /w constR & 0.303 & - \\
% \botline
% \end{tabular}
% \end{center}
% \end{table}
\begin{table}[h]
\caption{Time mean RMSEs of one-step CGDA and multi-step CGDA with different $\beta$.}\label{tab:rmse_tracer}
\begin{center}
\begin{tabular}{ccc}
\topline
 & $\beta=22$ & $\beta=111$ \\
\midline
One-step CGDA & 0.400 & 0.137  \\
Multi-step CGDA & 0.291 & 0.113 \\
\botline
\end{tabular}
\end{center}
\end{table}

\subsection{Sensitivity to data assimilation parameters}
\begin{figure}[htbp!]
 \centerline{\includegraphics[width=27pc]{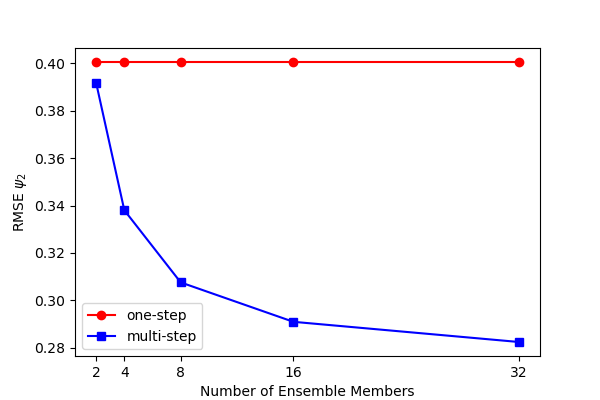}}
  \caption{Time mean posterior RMSEs of one-step CGDA and multi-step CGDA with different ensemble sizes.}\label{fig:ens}
\end{figure}

Figure \ref{fig:ens} shows the posterior mean RMSEs with different ensemble sizes. For the default configuration, the multi-step CGDA with only two ensemble members achieves smaller posterior errors than the one-step CGDA. This demonstrates the benefits of the multi-step CGDA that incorporates nonlinearities of the flow model. As the ensemble size increases, the posterior errors of multi-step CGDA rapidly decrease at first and then gradually level off. One can select a proper ensemble size to balance the trade-off between assimilation accuracy and computational cost.

\begin{figure}[htbp!]
 \centerline{\includegraphics[width=27pc]{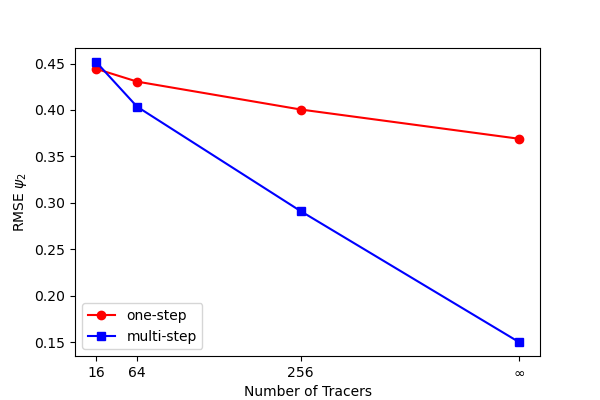}}
  \caption{Time mean posterior RMSEs of one-step CGDA and multi-step CGDA with different tracer numbers. The infinite-tracers case is solved conceptually with the upper-layer flow being directly and fully observed.}\label{fig:tracer_number}
\end{figure}

Figure \ref{fig:tracer_number} presents the posterior mean RMSEs varying with the number of tracers. The infinite-tracers case is solved conceptually by assuming that the upper-layer flow is directly and fully observed (see appendix C for details). As the number of tracers increases, both data assimilation methods produce lower posterior errors, with the multi-step CGDA achieving a faster error reduction rate than the one-step CGDA. Thus, the benefits of multi-step CGDA become more significant as the number of tracers increases.
% \begin{table}[h]
% \caption{Mean RMSEs for one-step CGDA and multi-step CGDA with different tracer numbers $L$.}\label{tab:tracer_number}
% \begin{center}
% \begin{tabular}{cccc}
% \topline
%  & $L=16$ & $L=64$ & $L=256$ \\
% \midline
% one-step CGDA & & 0.428 & 0.398  \\
% multi-step CGDA & & 0.400 & 0.288 \\
% \botline
% \end{tabular}
% \end{center}
% \end{table}

\begin{figure}[htbp!]
 \centerline{\includegraphics[width=33pc]{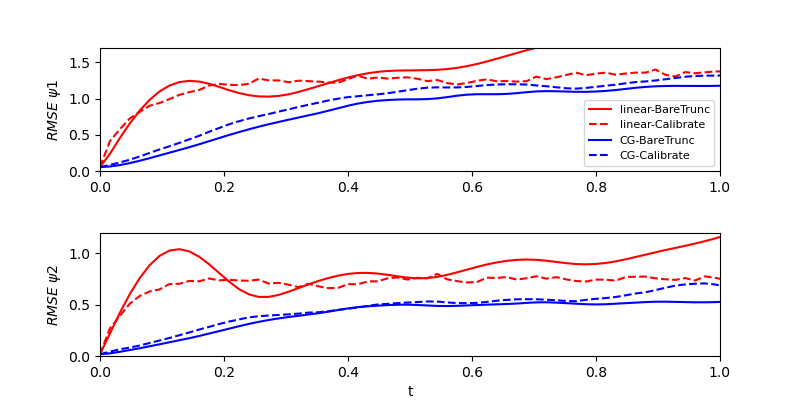}}
  \caption{RMSE series of the two-layer flow model free forecasts. Red solid lines indicate a bare-truncated linear flow model. Red dashed lines indicate a calibrated linear stochastic flow model. Blue solid lines indicate a bare-truncated conditional Gaussian nonlinear flow model. Blue dashed lines indicate a calibrated conditional Gaussian nonlinear stochastic flow model. The upper row corresponds to the upper layer. The lower row corresponds to the lower layer.}\label{fig:rmses_free}
\end{figure}

Figure \ref{fig:rmses_free} shows the time mean RMSEs of free forecasts given by the bare-truncated linear flow model (\ref{eq:linearqg}), the calibrated linear stochastic flow model (\ref{eq:lsm_eg}), the bare-truncated conditional Gaussian nonlinear flow model (\ref{eq:cg_raw}), and the calibrated conditional Gaussian nonlinear stochastic flow model (\ref{eq:cg_2layer}). Whether bare-truncated or calibrated, the conditional Gaussian nonlinear flow models exhibit slower error-growing rates than the linear flow models in short-term forecasts. For data assimilation with frequent observations, the short-term forecast performance is indicative enough for model evaluation. Therefore, the benefits of preserving additional nonlinearities in multi-step CGDA are further validated.

\subsection{Computational cost}
Finally, we summarize the computational cost of the multi-step CGDA focusing on recovering the lower layer from upper-layer pseudo-observations using a nonlinear conditional Gaussian flow model. The initial one-step CGDA that recovers surface flow from tracers can be substituted by other methods, which is not our focus here. Specifically, a single step of the CGDA using the nonlinear flow model has three computationally intensive parts: the backward sampling (\ref{eq:backsample}), the assembling of coefficient matrices (\ref{eq:A0a0}) and (\ref{eq:a1A1}), and the update of posterior mean and variance using (\ref{eq:cg_meanvar}).

The computational cost of multi-step CGDA algorithm largely depends on three input sizes: the assimilation steps $N_t$, the ensemble size $N_s$, and the cardinality of truncated Fourier modes set $|\mathcal{K}_{\mathrm{tr}}|$, where $\mathcal{K}_{\mathrm{tr}} \subseteq \mathcal{K}$. Consider a single assimilation step with one ensemble member; the cost of backward sampling is led by matrix multiplication and inversion, leading to an asymptotic complexity of $O(|\mathcal{K}_{\mathrm{tr}}|^2)$. The assembling of coefficient matrices involves calculating the linear terms, assigning the nonlinear terms, and aggregating linear and nonlinear terms, resulting in a cost of $O(|\mathcal{K}_{\mathrm{tr}}|^2)$. The update of mean and covariance cost $O(|\mathcal{K}_{\mathrm{tr}}|^3)$, with the cost led by matrix multiplication. Therefore, the total cost of a single step of CGDA using a nonlinear flow model is $O(N_t\times N_s\times|\mathcal{K}_{\mathrm{tr}}|^3)$ in the worst case.

\begin{figure}[htbp!]
 \centerline{\includegraphics[width=24pc]{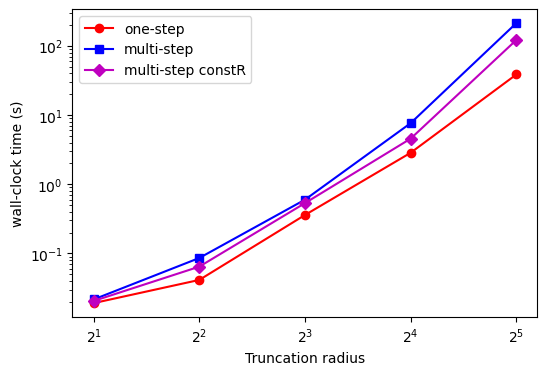}}
  \caption{The execution time (wall-clock time) of one-step CGDA with linear flow model (red), multi-step CGDA (blue), multi-step CGDA with constant covariance matrix (purple) varying with the Fourier modes truncation radius $r$. The square of truncation radius $r^2$ is proportional to $|\mathcal{K}_{\mathrm{tr}}|$. Both axes are displayed in log scale in order to reveal the linear relation.  The data assimilation methods are run for 10 assimilation steps. The multi-step data assimilation methods have 2 ensemble members.}\label{fig:compcost_r}
\end{figure}

\begin{figure}[htbp!]
 \centerline{\includegraphics[width=24pc]{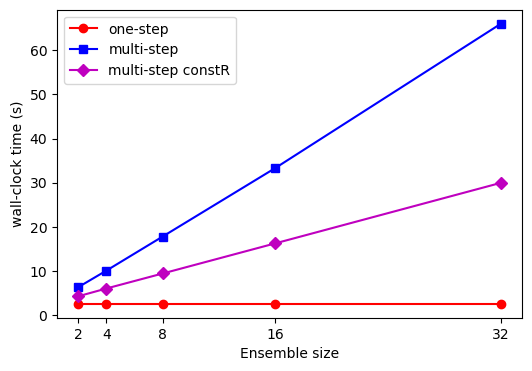}}
  \caption{The same as Fig. \ref{fig:compcost_r}, except for varying with ensemble size $N_s$. The truncation radius is fixed to $r=16$.}\label{fig:compcost_ens}
\end{figure}

However, it is possible to reduce the computational cost in practice. For backward sampling, leveraging the diagonal property of matrices $\bm{\mathsf{a}}$, $\bm{\mathsf{\Sigma}}_{\mathrm{E}}$, and $\bm{\mathsf{R}}_\mathrm{E}$ can reduce the cost to $O(|\mathcal{K}_{\mathrm{tr}}|)$. A constant covariance matrix can be used. Therefore, only the mean update is needed, which further reduces the computational costs. Figures \ref{fig:compcost_r} and \ref{fig:compcost_ens} show the execution time of the one-step CGDA with linear flow model, multi-step CGDA, and multi-step CGDA with constant covariance as a function of truncation radius and ensemble size, respectively. Using a constant covariance matrix nearly cuts the cost by half, as shown in Fig. \ref{fig:compcost_ens}. In the QG application, using a constant covariance achieves a mean posterior RMSE of 0.303, which is close to the multi-step CGDA's 0.291 and significantly lower than the one-step CGDA's 0.400.

\section{Conclusion and final discussions}
A nonlinear closed-form data assimilation scheme is developed for multi-layer flow fields with surface observations. The multi-step CGDA accounts for the nonlinearity in both observations and flow, with analytic formulae to solve the posterior mean and covariance. This is achieved by sequentially filtering each layer from top to bottom. The multi-step CGDA is tested in the two-layer QG model compared to the one-step CGDA with linear flow models. The multi-step CGDA produces smaller posterior RMSE than the one-step CGDA and can capture non-Gaussian posteriors. Consistent results are found for different turbulent regimes of the two-layer QG model, with the benefits of multi-step CGDA amplified as the tracer number and ensemble size increase. In particular, using a constant covariance can greatly reduce computational costs while maintaining relatively good filter performance in certain cases.

Compared to the OI method that linearly connects corrections across layers \citep{molcard_lagrangian_2005}, the one-step CGDA with linear stochastic flow models connects layers via eigenvectors. These two methods are equivalent when both the observation and flow are linear Gaussian processes. However, in Lagrangian data assimilation, the observation process is inherently nonlinear.  Although the OI method can use a nonlinear flow model to provide the prior, the Gaussian assumption does not hold. In comparison, the one-step CGDA allows a nonlinear observation process with the nonlinearity exactly considered in the formulation. Furthermore, the multi-step CGDA can handle flow model nonlinearity in its subsequent steps, thus better propagating the corrections across layers. Future studies could apply the multi-step CGDA to flow fields with more than two layers to validate this.

The backward smoother sampling formula is used to generate pseudo-observations of the upper layer. This leads to an unbiased estimation but is impractical in the online filtering setting. An optimal filter-based forward sampling formula is needed to test the conclusions in filtering practice. For scenarios where a smoother is applicable, the smoother version of the multi-step CGDA can be utilized for parameter estimation. For instance, constant forcing parameters, such as topography, can be estimated using the multi-step conditional Gaussian smoother. Potential applications include recovering the ocean bottom topography based on surface tracer observations and the associated uncertainty quantification.

The assimilation of each layer is modularized in the multi-step data assimilation, meaning it is a general framework. One can switch to alternative data assimilation schemes at any step, not necessarily restricted to CGDA. For example, a more advanced data assimilation scheme than the one-step CGDA could be used at the first data assimilation step to improve the recovery of the surface flow. This improvement will propagate downward and lead to a better recovery of the lower layers as well. In addition, the multi-step CGDA is natural for, but not restricted to, Lagrangian data. Eulerian data is adoptable if its governing equation is provided and the data is frequent enough. An example work that combines Eulerian and Lagrangian data assimilation under the CGDA framework is presented in \cite{deng_lemda_2024}. Besides, covariance inflation \citep{anderson_monte_1999} can be applied to the pseudo-observations, which may help compensate for the approximation errors described in Section \ref{sec:2c} and further improve the multi-step data assimilation accuracy.

%%%%%%%%%%%%%%%%%%%%%%%%%%%%%%%%%%%%%%%%%%%%%%%%%%%%%%%%%%%%%%%%%%%%%
% TABLES---INSERT NEAR IN-TEXT DISCUSSION
%%%%%%%%%%%%%%%%%%%%%%%%%%%%%%%%%%%%%%%%%%%%%%%%%%%%%%%%%%%%%%%%%%%%%
%% Enter tables near where they are discussed within the document.
%%
%
%\begin{table}[h]
%\caption{This is a sample table caption and table layout.
%Table from Lorenz (1963).}\label{t1}
%\begin{center}
%\begin{tabular}{ccccrrcrc}
%\topline
%$N$ & $X$ & $Y$ & $Z$\\
%\midline
% 0000 & 0000 & 0010 & 0000 \\
% 0005 & 0004 & 0012 & 0000 \\
% 0010 & 0009 & 0020 & 0000 \\
% 0015 & 0016 & 0036 & 0002 \\
% 0020 & 0030 & 0066 & 0007 \\
% 0025 & 0054 & 0115 & 0024 \\
%\botline
%\end{tabular}
%\end{center}
%\end{table}

%%%%%%%%%%%%%%%%%%%%%%%%%%%%%%%%%%%%%%%%%%%%%%%%%%%%%%%%%%%%%%%%%%%%%
% FIGURES---INSERT NEAR IN-TEXT DISCUSSION
%%%%%%%%%%%%%%%%%%%%%%%%%%%%%%%%%%%%%%%%%%%%%%%%%%%%%%%%%%%%%%%%%%%%%
%%  Enter figures near where they are discussed within the document.
%%
%
%\begin{figure}[t]
%  \noindent\includegraphics[width=19pc,angle=0]{figure01.pdf}\\
%  \caption{Enter the caption for your figure here.  Repeat as
%  necessary for each of your figures. Figure from \protect\cite{Knutti2008}.}\label{f1}
%\end{figure}

\clearpage
%%%%%%%%%%%%%%%%%%%%%%%%%%%%%%%%%%%%%%%%%%%%%%%%%%%%%%%%%%%%%%%%%%%%%
% ACKNOWLEDGMENTS
%%%%%%%%%%%%%%%%%%%%%%%%%%%%%%%%%%%%%%%%%%%%%%%%%%%%%%%%%%%%%%%%%%%%%
\acknowledgments
The research of N.C. is funded by the Office of Naval Research N00014-24-1-2244 and the Army Research Office W911NF-23-1-0118. Z.W. is partially supported as a research assistant under the first grant. The research of D.Q. is partially funded 
 by the Office of Naval Research N00014-24-1-2192 and the National Science Foundation DMS-2407361 and OAC-2232872.
 
%%%%%%%%%%%%%%%%%%%%%%%%%%%%%%%%%%%%%%%%%%%%%%%%%%%%%%%%%%%%%%%%%%%%%
% DATA AVAILABILITY STATEMENT
%%%%%%%%%%%%%%%%%%%%%%%%%%%%%%%%%%%%%%%%%%%%%%%%%%%%%%%%%%%%%%%%%%%%%
%
%
\datastatement
The code can be found at:
\url{https://github.com/zhongruiw/MultistepCGDA}.

%%%%%%%%%%%%%%%%%%%%%%%%%%%%%%%%%%%%%%%%%%%%%%%%%%%%%%%%%%%%%%%%%%%%%
% APPENDIXES
%%%%%%%%%%%%%%%%%%%%%%%%%%%%%%%%%%%%%%%%%%%%%%%%%%%%%%%%%%%%%%%%%%%%%

% \appendix[A]
\section*{Appendix A: Calibration of Stochastic Models}\label{appendixA}

\subsection{Calibrating LSMs}\label{Appendix:Aa}
The parameters $\gamma$, $\omega$, $f$, and $\sigma$ of the complex OU process can be determined by matching the four statistics computed from the observational time series. These statistics are the mean $m$, the variance $V$, and the real and imaginary parts of the decorrelation time, $T$ and $\theta$. The latter two can be calculated as follows:
\begin{equation}
R(s) = \frac{E\left[(\hat{\psi}_{
t} - \overline{\hat{\psi}}_{
\infty})(\hat{\psi}_{
t+s} - \overline{\hat{\psi}}_{
\infty})\right]}{\mathrm{var}(\hat{\psi}_{
\infty})}. \quad
\end{equation}
\begin{equation}
\int_0^{\infty} R(\tau) d\tau = T - i\theta,
\end{equation}
where \(R(s)\) is the autocorrelation function (ACF). Then the four parameters are uniquely determined by these four statistical quantities via the following formulae
\begin{equation}
\gamma = \frac{T}{T^2 + \theta^2}, \quad \omega = \frac{\theta}{T^2 + \theta^2}, \quad f = \frac{m(T - i\theta)}{T^2 + \theta^2}, \quad \text{and} \quad \sigma = \sqrt{\frac{2VT}{T^2 + \theta^2}}.
\end{equation}
In practice, an infinite time series is unavailable. Using a truncated time series introduces errors in the estimation of ACF and decorrelation time. Therefore, we choose an alternative approach to estimate $d$ and $\omega$. By using an ansatz of the ACF, where the autocorrelation of the real part of $\hat{\psi}$ equals $e^{-\gamma t}cos\omega t$, and the cross-correlation between the real and imaginary parts equals $e^{-\gamma t}sin\omega t$ \citep{majda_stochastic_2012}, we can fit the ansatz to the observations thus getting $\gamma$ and $\omega$. Note that when using ACF for fitting, the signs of $\omega$ need to be calibrated according to the ones based on CCF. Because the cosine function is even.

\subsection{Calibrating conditional Gaussian stochastic models}\label{Appendix:Ab}
Assuming the parameters of QG equations are known, we only calibrate the noise strength $\bm{\mathsf{\Sigma}}_1$ and $\bm{\mathsf{\Sigma}}_2$ in the conditional Gaussian model (\ref{eq:cg_2layer}):
\begin{equation}
    \sigma_{j,\mathbf{k}}= \sqrt{\frac{\mathrm{var}\left(\hat{\epsilon}_{j,\mathbf{k}}(t\to \infty) \right)}{2} },\ j=1,2
\end{equation}
where $\sigma_{j,\mathbf{k}}$ is the diagonal entry of $\bm{\mathsf{{\Sigma}}}_j$ corresponding to wave number $\mathbf{k}$. $\hat{\epsilon}_{j,\mathbf{k}}(t\to \infty)$ is the residual at each time step for the equilibrium state.

\subsection{Evaluation metrics}\label{Appendix:Ac}
Several metrics are used for evaluating the model calibration. The energy of $\psi_{\mathbf{k}}$ is defined as
\begin{equation}
    \mathrm{E}_{\psi_{\mathbf{k}}}=|\psi_{\mathbf{k}}|^2.
\end{equation}
The kinetic energy (KE) of the two-layer flow field corresponding to a wave number $\mathbf{k}$ is
\begin{equation}
    \mathrm{KE}_\mathbf{k} = \frac{1}{2}(|\nabla\psi_{1,\mathbf{k}}|^2+|\nabla\psi_{2,\mathbf{k}}|^2) = \frac{\mathbf{k}^2}{2}(|\psi_{1,\mathbf{k}}|^2+|\psi_{2,\mathbf{k}}|^2).
\end{equation}
The available potential energy (APE) for wave number $\mathbf{k}$ is
\begin{equation}
    \mathrm{APE}_{\mathbf{k}}=\frac{k_d^2}{4}|\psi_{1,\mathbf{k}}-\psi_{2,\mathbf{k}}|^2.
\end{equation}
The total energy (E) of the two-layer flow for wave number $\mathbf{k}$ is
\begin{equation}
    \mathrm{E}_\mathbf{k} = \mathrm{KE}_\mathbf{k}+\mathrm{APE}_{\mathbf{k}}.
\end{equation}
The enstrophy ($\mathcal{E}$) of the two-layer flow for wave number $\mathbf{k}$ is
\begin{equation}
    \mathcal{E}_\mathbf{k} = \frac{1}{2}(|q_{1,\mathbf{k}}|^2+|q_{2,\mathbf{k}}|^2).
\end{equation}
Note that all the energy in Fourier space is normalized by $1/{|\mathcal{K}|^2}$ in evaluation. The energy for each Fourier mode is then distributed into discrete bins of $k=0,1,2,...$ according to the magnitude $|\mathbf{k}|$. For example,
\begin{equation}
    \mathrm{E}_k = \sum^{\mathcal{K}}_{\substack{\mathbf{k} \\ \lfloor|\mathbf{k}|\rfloor = k}} (\lceil|\mathbf{k}|\rceil-|\mathbf{k}|)\mathrm{E}_\mathbf{k} +
    \sum^{\mathcal{K}}_{\substack{\mathbf{k} \\ \lceil|\mathbf{k}|\rceil = k}} (|\mathbf{k}|-\lfloor|\mathbf{k}|\rfloor)\mathrm{E}_\mathbf{k},
\end{equation}
where $\lfloor\cdot\rfloor$ and $\lceil\cdot\rceil$ are the floor and ceiling functions, respectively.

\subsection{Conversion between eigenmodes energy and layers energy}\label{Appendix:Ad}
Suppose $\psi_{1,\mathbf{k}}= r_1\psi_{\mathrm{E}1,\mathbf{k}} + r_2 \psi_{\mathrm{E}2,\mathbf{k}}$ is a linear combination of $\psi_{\mathrm{E}1,\mathbf{k}}$ and $\psi_{\mathrm{E}2,\mathbf{k}}$
.Omitting $\mathbf{k}$ for convenience, the energy of $\psi_{1,\mathbf{k}}$ is
\begin{align*}
       \mathrm{E}_{{\psi_1}} &= |r_1\psi_{\mathrm{E}1} + r_2 \psi_{\mathrm{E}2}|^2 \\
       &= [r_1\mathrm{Re}(\psi_{\mathrm{E}1}) + r_2\mathrm{Re}(\psi_{\mathrm{E}2})]^2 + [r_1\mathrm{Im}(\psi_{\mathrm{E}1}) + r_2\mathrm{Im}(\psi_{\mathrm{E}2})]^2 \\
       &= r_1^2|\psi_{\mathrm{E}1}|^2 + r_2^2|\psi_{\mathrm{E}2}|^2 + 2r_1r_2[\mathrm{Re}(\psi_{\mathrm{E}1})\mathrm{Re}(\psi_{\mathrm{E}2}) + \mathrm{Im}(\psi_{\mathrm{E}1})\mathrm{Im}(\psi_{\mathrm{E}2})] \\
       &= r_1^2\mathrm{E}_{\mathrm{E}1_\psi} + r_2^2\mathrm{E}_{\mathrm{E}1_{\mathrm{E}2}} + 2r_1r_2[\mathrm{Re}(\psi_{\mathrm{E}1})\mathrm{Re}(\psi_{\mathrm{E}2}) + \mathrm{Im}(\psi_{\mathrm{E}1})\mathrm{Im}(\psi_{\mathrm{E}2})].
\end{align*}
A cross term $2r_1r_2[\mathrm{Re}(\psi_{\mathrm{E}1})\mathrm{Re}(\psi_{\mathrm{E}2}) + \mathrm{Im}(\psi_{\mathrm{E}1})\mathrm{Im}(\psi_{\mathrm{E}2})]$ emerges in energy after the linear transformation from eigenmodes to layers. This means, an exact match of $\mathrm{E}_{\psi_{\mathrm{E}1}}, \mathrm{E}_{\psi_{\mathrm{E}2}}$ does not necessarily assure a perfect match of $\mathrm{E}_{\psi_1}$. The cross term can be explicitly considered in the linear stochastic modeling by matching the statistics of $\mathrm{Re}(\psi_{\mathrm{E}1})\mathrm{Re}(\psi_{\mathrm{E}2})$ and $\mathrm{Im}(\psi_{\mathrm{E}1})\mathrm{Im}(\psi_{\mathrm{E}2})$.

\subsection{Results on two-layer QG system}\label{Appendix:Ae}
\begin{figure}[h]
 \centerline{\includegraphics[width=33pc]{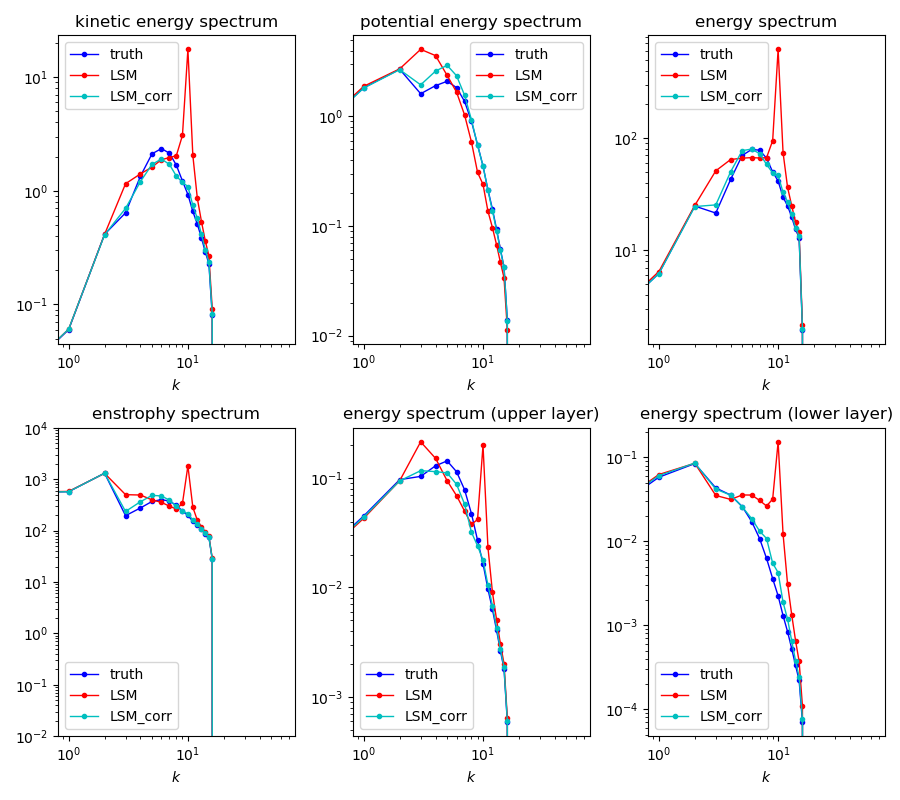}}
  \caption{The energy spectrum of two-layer modes: true signals (blue), LSMs (red), and LSMs with correlated noises (cyan).}\label{fig:spectrum_layer_beta22}
\end{figure}

To evaluate the calibration, we can compare the statistics, including mean, variance, and energy metrics, of the LSMs to the true signals from QG. We can also plot the trajectories to get an intuitive comparison. The calibrated LSMs of the eigenmodes match well for the eigenmodes. However, after the transformation to two-layer quantities,  only the mean is matched. Mismatches in variance and energy metrics exist for two-layer quantities but can be eliminated by allowing correlated noises of eigenmodes (see appendix B for details of the LSMs with correlated noises and Fig. \ref{fig:spectrum_layer_beta22} for energy spectrum results).

\section*{Appendix B: Linear Stochastic Models with Correlated Noises}\label{appdenixB}

\subsection{Model form}
To consider the cross terms of two eigenmodes, we let the LSMs have correlated noises:
\begin{equation}
\begin{aligned}
    \frac{\mathrm{d}}{\mathrm{d}t}\begin{bmatrix}
       \mathrm{Re}(\hat{\psi}_{\mathrm{E}1}) \\
       i\mathrm{Im}(\hat{\psi}_{\mathrm{E}1}) \\
       \mathrm{Re}(\hat{\psi}_{\mathrm{E}2}) \\
       i\mathrm{Im}(\hat{\psi}_{\mathrm{E}2})
    \end{bmatrix}
    =
    \begin{bmatrix}
       -\gamma_{1}, i\omega_{1}, 0, 0\\
        i\omega_{1}, -\gamma_{1}, 0, 0 \\
        0, 0, -\gamma_{2}, i\omega_{2} \\
        0, 0 , i\omega_{1},-\gamma_{1}
    \end{bmatrix}
    \begin{bmatrix}
       \mathrm{Re}(\hat{\psi}_{\mathrm{E}1}) \\
       i\mathrm{Im}(\hat{\psi}_{\mathrm{E}1}) \\
       \mathrm{Re}(\hat{\psi}_{\mathrm{E}2}) \\
       i\mathrm{Im}(\hat{\psi}_{\mathrm{E}2})
    \end{bmatrix}  +
    \begin{bmatrix}
       \mathrm{Re}(f_{1}) \\
       i\mathrm{Im}(f_{1}) \\
       \mathrm{Re}(f_{2}) \\
       i\mathrm{Im}(f_{2})
    \end{bmatrix}
     +
    \begin{bmatrix}
       \sigma_{1}, 0, c_{\mathrm{R}}, 0\\
        0, \sigma_{1}, 0, c_{\mathrm{I}} \\
        c_{\mathrm{R}}, 0, \sigma_{2}, 0 \\
        0, c_{\mathrm{I}}, 0 , \sigma_{2}
    \end{bmatrix}
    \begin{bmatrix}
       \mathrm{Re}(\dot{W}_{\mathrm{E}1}) \\
       i\mathrm{Im}(\dot{W}_{\mathrm{E}1}) \\
       \mathrm{Re}(\dot{W}_{\mathrm{E}2}) \\
       i\mathrm{Im}(\dot{W}_{\mathrm{E}2})
    \end{bmatrix}
\end{aligned}
\end{equation}
The subscript $\mathbf{k}$ denoting wavenumber is omitted here for convenience. Note that the noise coefficient matrix is no longer diagonal, thus allowing correlation in noise.

\subsection{Calibration}
For LSMs with correlated noise, the way to calibrate parameters $\gamma$, $\omega$, and $f$ remains the same as in appendix \ref{Appendix:Aa}. Because  $\gamma$, $\omega$, and $f$  only depend on the mean and decorrelation time. The calibration of $\sigma_{1}$, $\sigma_{2}$, $c_{\mathrm{R}}$, and $c_{\mathrm{I}}$ is a little more complicated, but can be analytically determined by matching $\mathrm{var}(\hat{\psi}_{\mathrm{E}1})$, $\mathrm{var}(\hat{\psi}_{\mathrm{E}2})$, $\mathrm{cov}(\mathrm{Re}(\hat{\psi}_{\mathrm{E}1})\mathrm{Re}(\hat{\psi}_{\mathrm{E}2}))$, and $\mathrm{cov}(\mathrm{Im}(\hat{\psi}_{\mathrm{E}1})\mathrm{Im}(\hat{\psi}_{\mathrm{E}2}))$.
\begin{equation}\label{eq:Acov}
    \begin{aligned}
        \mathrm{var}(\hat{\psi}_{\mathrm{E}1}) &= \frac{2\sigma_{1}^2 + c_{\mathrm{R}}^2 + c_{\mathrm{I}}^2}{2\gamma_{1}}, \\
        \mathrm{var}(\hat{\psi}_{\mathrm{E}2}) &= \frac{2\sigma_{2}^2 + c_{\mathrm{R}}^2 + c_{\mathrm{I}}^2}{2\gamma_{2}}, \\ \mathrm{cov}(\mathrm{Re}(\hat{\psi}_{\mathrm{E}1})\mathrm{Re}(\hat{\psi}_{\mathrm{E}2})) &= \frac{c_{\mathrm{R}}(\sigma_{1} + \sigma_{2})}{\gamma_{1} + \gamma_{2}},  \\\mathrm{cov}(\mathrm{Im}(\hat{\psi}_{\mathrm{E}1})\mathrm{Im}(\hat{\psi}_{\mathrm{E}2})) &= \frac{c_{\mathrm{I}}(\sigma_{1} + \sigma_{2})}{\gamma_{1} + \gamma_{2}}.
    \end{aligned}
\end{equation}
One can derive (\ref{eq:Acov}) using the general Itô Formula \citep{oksendal_ito_2003}. The solution to this system of nonlinear equations is not unique. But we can rule out some solutions by enforcing certain regularizations, for example, $\sigma_{1}\geq 0$ and $\sigma_{2}\geq 0$. At last, only one solution is adopted:
\begin{equation}
    \begin{aligned}
        \sigma_{1} &= \sqrt{2}b_3 \sqrt{\frac{b_1+b_2-2\sqrt{b_1b_2-2(b_3^2+b_4^2)}}{(b_1-b_2)^2 + 8(b_3^2  + b_4^2)}},   \\
        \sigma_{2} &= \sqrt{2}b_4 \sqrt{\frac{b_1+b_2-2\sqrt{b_1b_2-2(b_3^2+b_4^2)}}{(b_1-b_2)^2 + 8(b_3^2  + b_4^2)}}, \\
        c_{\mathrm{R}} &= \frac{\sqrt{2}}{2} \left(b_1+\sqrt{b_1b_2-2(b_3^2+b_4^2)}\right) \sqrt{\frac{b_1+b_2-2\sqrt{b_1b_2-2(b_3^2+b_4^2)}}{(b_1-b_2)^2 + 8(b_3^2  + b_4^2)}}, \\
        c_{\mathrm{I}} &= \frac{\sqrt{2}}{2} \left(b_2+\sqrt{b_1b_2-2(b_3^2+b_4^2)}\right) \sqrt{\frac{b_1+b_2-2\sqrt{b_1b_2-2(b_3^2+b_4^2)}}{(b_1-b_2)^2 + 8(b_3^2  + b_4^2)}},
    \end{aligned}
\end{equation}
where $b_1=2\gamma_{1}\mathrm{var}(\hat{\psi}_{\mathrm{E}1})$, $b_2=2\gamma_{2}\mathrm{var}(\hat{\psi}_{\mathrm{E}2})$, $b_3=(\gamma_{1} + \gamma_{2})\mathrm{cov}(\mathrm{Re}(\hat{\psi}_{\mathrm{E}1})\mathrm{Re}(\hat{\psi}_{\mathrm{E}2}))$, and $b_4=(\gamma_{1} + \gamma_{2})\mathrm{cov}(\mathrm{Im}(\hat{\psi}_{\mathrm{E}1})\mathrm{Im}(\hat{\psi}_{\mathrm{E}2}))$.

\subsection{CGDA}
For LSMs of eigenmodes with correlated noise, the CGDA formulae remain the same as (\ref{eq:lsmeg_da}). But the size of noise coefficient matrix $\bm{\mathsf{\Sigma}}_{\mathrm{E}}$ expands to $4|\mathcal{K}| \times 4|\mathcal{K}|$ if $c_{\mathrm{R}} \neq c_{\mathrm{I}}$, because the real and imaginary parts of $\boldsymbol{\Psi}_\mathrm{E}$ needs to be separated. Fortunately, we observe that $c_{\mathrm{R}}$ is close to $c_{\mathrm{I}}$ in the QG case. So by letting $c=(c_{\mathrm{R}} + c_{\mathrm{I}})/2$ and replace $c_{\mathrm{R}}$ and $c_{\mathrm{I}}$ by $c$, the size of $\bm{\mathsf{\Sigma}}_{\mathrm{E}}$ can remain unchanged. Only a small modification needs to be made:
\begin{equation}
    \bm{\mathsf{\Sigma}}_{\mathrm{E}}
    =\sqrt{2}\begin{bmatrix}
       \ddots &  &   & \ddots &  &  \\
         & \sigma_{1,\mathbf{k}} &  & &c_{\mathbf{k}} &  \\
          &  & \ddots &  & & \ddots \\
         \ddots &  &   & \ddots &  &  \\
         & c_{\mathbf{k}} &  & &\sigma_{2,\mathbf{k}} &  \\
          &  & \ddots &  & & \ddots
    \end{bmatrix}.
\end{equation}
With proper permutations, $\bm{\mathsf{\Sigma}}_{\mathrm{E}}$ can become a block-diagonal matrix.

\begin{table}[h]
\begin{center}
\begin{tabular}{cc}
\topline
one-step CGDA &  0.400  \\
one-step CGDA w/ correlated noises &  0.372  \\
multi-step CGDA & 0.291 \\
\botline
\end{tabular}
\end{center}
\caption{Time mean RMSEs for one-step CGDA, one-step CGDA using LSMs with correlated noises, and multi-step CGDA.}\label{tab:rmse_lsmcn}
\end{table}

\section*{Appendix C: Upper Layer Fully Observed Case}\label{appendixC}
\subsection{CGDA with linear stochastic flow models}
\label{appendix:lsm_flow}
When the upper-layer flow is directly and fully observed, the two-layer linear stochastic flow model itself becomes conditional Gaussian. From the eigenmodes model (\ref{eq:ou}) and the conversion formulae between two-layer modes and eigenmodes, the two-layer linear stochastic flow model can be derived as
\begin{equation}\label{eq:lsm_layer}
\begin{aligned}
\frac{\mathrm{d}\hat \psi_{1,\mathbf{k}}}{\mathrm{d}t} &=  (-\Gamma_{1,\mathbf{k}} + i\Omega_{1,\mathbf{k}}) \hat \psi_{1,\mathbf{k}} + (-\Gamma_{2,\mathbf{k}} + i\Omega_{2,\mathbf{k}}) \hat \psi_{2,\mathbf{k}} + f_{1,\mathbf{k}}  + \Sigma_{1,\mathbf{k}} \dot{W}_{\mathrm{E}1,\mathbf{k}}+ \Sigma_{2,\mathbf{k}} \dot{W}_{\mathrm{E}2,\mathbf{k}}, \\
\frac{\mathrm{d}\hat \psi_{2,\mathbf{k}}}{\mathrm{d}t} &= \left( (-\gamma_{1,\mathbf{k}} + i\omega_{1,\mathbf{k}}) \hat \psi_{1,\mathbf{k}} + (-\gamma_{2,\mathbf{k}} + i\omega_{2,\mathbf{k}}) \hat \psi_{2,\mathbf{k}} + f_{2,\mathbf{k}} \right)\mathrm{d} + \sigma_{1,\mathbf{k}} \dot{W}_{\mathrm{E}1,\mathbf{k}}+ \sigma_{2,\mathbf{k}} \dot{W}_{\mathrm{E}2,\mathbf{k}}
\end{aligned}
\end{equation}
where
\begin{equation}
    \begin{aligned}
        -\Gamma_{1,\mathbf{k}} + i\Omega_{1,\mathbf{k}}&= \frac{r^{(1)}_{1,\mathbf{k}}r^{(2)}_{2,\mathbf{k}}(-\gamma_{\mathrm{E}1,\mathbf{k}} + i\omega_{1,\mathbf{k}} ) - r^{(2)}_{1,\mathbf{k}}r^{(1)}_{2,\mathbf{k}}(-\gamma_{\mathrm{E}2,\mathbf{k}}+ \omega_{2,\mathbf{k}})} {r^{(1)}_{1,\mathbf{k}}r^{(2)}_{2,\mathbf{k}} - r^{(2)}_{1,\mathbf{k}}r^{(1)}_{2,\mathbf{k}}}, \\
        -\Gamma_{2,\mathbf{k}}+i\Omega_{2,\mathbf{k}} &= \frac{r^{(1)}_{1,\mathbf{k}}r^{(1)}_{2,\mathbf{k}}\left(-(-\gamma_{\mathrm{E}1,\mathbf{k}}+i\omega_{1,\mathbf{k}}) + (-\gamma_{\mathrm{E}2,\mathbf{k}}+ i\omega_{2,\mathbf{k}})\right)}{r^{(1)}_{1,\mathbf{k}}r^{(2)}_{2,\mathbf{k}} - r^{(2)}_{1,\mathbf{k}}r^{(1)}_{2,\mathbf{k}}},  \\
        -\gamma_{1,\mathbf{k}}+ i\omega_{1,\mathbf{k}} &= \frac{r^{(2)}_{1,\mathbf{k}}r^{(2)}_{2,\mathbf{k}}\left((-\gamma_{\mathrm{E}1,\mathbf{k}}+i\omega_{1,\mathbf{k}}) -(-\gamma_{\mathrm{E}2,\mathbf{k}}+i\omega_{2,\mathbf{k}})\right)}{r^{(1)}_{1,\mathbf{k}}r^{(2)}_{2,\mathbf{k}} - r^{(2)}_{1,\mathbf{k}}r^{(1)}_{2,\mathbf{k}}},
        \\
        -\gamma_{2,\mathbf{k}} +i\omega_{2,\mathbf{k}} &= \frac{-r^{(2)}_{1,\mathbf{k}}r^{(1)}_{2,\mathbf{k}}(-\gamma_{\mathrm{E}1,\mathbf{k}}+i\omega_{1,\mathbf{k}}) + r^{(1)}_{1,\mathbf{k}}r^{(2)}_{2,\mathbf{k}}(-\gamma_{\mathrm{E}2,\mathbf{k}}+i\omega_{2,\mathbf{k}})}{r^{(1)}_{1,\mathbf{k}}r^{(2)}_{2,\mathbf{k}} - r^{(2)}_{1,\mathbf{k}}r^{(1)}_{2,\mathbf{k}}},
        \\
        f_{1,\mathbf{k}} &= r^{(1)}_{1,\mathbf{k}} f_{\mathrm{E}1,\mathbf{k}} + r^{(1)}_{2,\mathbf{k}}f_{\mathrm{E}2,\mathbf{k}},
        \ \ \ f_{2,\mathbf{k}} = r^{(2)}_{1,\mathbf{k}} f_{\mathrm{E}1,\mathbf{k}} + r^{(2)}_{2,\mathbf{k}}f_{\mathrm{E}2,\mathbf{k}},
        \\
        \Sigma_{1,\mathbf{k}} &= r^{(1)}_{1,\mathbf{k}}\sigma_{1,\mathbf{k}}, \ \ \ \ \ \ \ \ \Sigma_{2,\mathbf{k}} = r^{(1)}_{2,\mathbf{k}}\sigma_{\mathrm{E}2,\mathbf{k}}, \\
                \sigma_{1,\mathbf{k}} &= r^{(2)}_{1,\mathbf{k}}\sigma_{1,\mathbf{k}},\ \ \ \ \ \ \ \   \sigma_{2,\mathbf{k}} = r^{(2)}_{2,\mathbf{k}}\sigma_{\mathrm{E}2,\mathbf{k}},
    \end{aligned}
\end{equation}
Write (\ref{eq:lsm_layer}) in matrix form,
\begin{equation}\label{eq:lsm_2layer}
    \begin{aligned}
\frac{\mathrm{d}\boldsymbol{\Psi}_1 }{\mathrm{d}t} &= \mathbf{A}_0(\boldsymbol{\Psi}_1,t) + \bm{\mathsf{A}}_1 \boldsymbol{\Psi}_2 + \bm{\mathsf{\Sigma}}_{1,1} \dot{\mathbf{W}}_{\mathrm{E}1}  + \bm{\mathsf{\Sigma}}_{1,2} \dot{\mathbf{W}}_{\mathrm{E}2}, \\
\frac{\mathrm{d}\boldsymbol{\Psi}_2 }{\mathrm{d}t} &= \mathbf{a}_0(\boldsymbol{\Psi}_1,t) + \bm{\mathsf{a}}_1 \boldsymbol{\Psi}_2 + \bm{\mathsf{\Sigma}}_{2,1} \dot{\mathbf{W}}_{\mathrm{E}1}+ \bm{\mathsf{\Sigma}}_{2,2} \dot{\mathbf{W}}_{\mathrm{E}2},
    \end{aligned}
\end{equation}
where $\mathbf{A}_0=(-\bm{\mathsf{\Gamma}}_{1,1} + i\bm{\mathsf{\Omega}}_{1,1})\boldsymbol{\Psi}_1 + \mathbf{F}_1$, $\bm{\mathsf{A}}_1=-\bm{\mathsf{\Gamma}}_{2,2} + i\bm{\mathsf{\Omega}}_{2,2}$, $\mathbf{a}_0=(-\bm{\mathsf{\Gamma}}_{2,1} + i\bm{\mathsf{\Omega}}_{2,1})\boldsymbol{\Psi}_1+ \mathbf{F}_2$, and $\bm{\mathsf{a}}_1=-\bm{\mathsf{\Gamma}}_{2,2} + i\bm{\mathsf{\Omega}}_{2,2}$. The damping $\bm{\mathsf{\Gamma}}_{i,j}$, phase $\bm{\mathsf{\Omega}}_{i,j}$, and
noise $\bm{\mathsf{\Sigma}}_{i,j}$ parameters are diagonal matrices. The forcing term $\mathbf{F}_i$ is a vector. The conditional Gaussian system (\ref{eq:lsm_2layer}) has the following formulae to solve posterior mean and covariance of $\boldsymbol{\Psi}_2$ given past trajectories of $\boldsymbol{\Psi}_1$:
\begin{subequations}
\begin{align}\label{eq:mean_lsmlayer}
        \frac{\mathrm{d}\Bar{\boldsymbol{\Psi}}_2}{\mathrm{d}t} &=  (\mathbf{a}_0 + \bm{\mathsf{a}}_1 \Bar{\boldsymbol{\Psi}}_2) + (\tilde{\bm{\mathsf{\Sigma}}}_2 \circ \tilde{\bm{\mathsf{\Sigma}}}_1 + \bm{\mathsf{R}}_2\bm{\mathsf{A}}_1^*) (\tilde{\bm{\mathsf{\Sigma}}}_1 \circ \tilde{\bm{\mathsf{\Sigma}}}_1)^{-1}
        \left(\frac{\mathrm{d}\boldsymbol{\Psi}_1}{\mathrm{d}t}-(\mathbf{A}_0 + \bm{\mathsf{A}}_1\Bar{\boldsymbol{\Psi}}_2)\right), \\
    \frac{\mathrm{d}\bm{\mathsf{R}}_2}{\mathrm{d}t} &= \bm{\mathsf{a}}_1\bm{\mathsf{R}}_2+\bm{\mathsf{R}}_2\bm{\mathsf{a}}_1^* + \tilde{\mathbf{{\bm{\mathsf{\Sigma}}}}}_2 \circ \tilde{\mathbf{{\bm{\mathsf{\Sigma}}}}}_2
    - (\tilde{\bm{\mathsf{\Sigma}}}_2 \circ \tilde{\bm{\mathsf{\Sigma}}}_1 + \bm{\mathsf{R}}_2\bm{\mathsf{A}}_1^*) (\tilde{\bm{\mathsf{\Sigma}}}_1 \circ \tilde{\bm{\mathsf{\Sigma}}}_1)^{-1} (\tilde{\bm{\mathsf{\Sigma}}}_1 \circ \tilde{\bm{\mathsf{\Sigma}}}_2 + \bm{\mathsf{A}}_1{\bm{\mathsf{R}}}_2^*),\label{eq:cov_lsmlayer}
\end{align}
\end{subequations}
where
\begin{equation*}
    \tilde{\bm{\mathsf{\Sigma}}}_i \circ \tilde{\bm{\mathsf{\Sigma}}}_j= {\bm{\mathsf{\Sigma}}}_{i,1} {\bm{\mathsf{\Sigma}}}_{j,1}^* + {\bm{\mathsf{\Sigma}}}_{i,2} {\bm{\mathsf{\Sigma}}}_{j,2}^*, \quad i=1,2;\ j=1,2
\end{equation*}

\subsection{CGDA with nonlinear stochastic flow models}
In the modified two-layer QG system (\ref{eq:cg_qg}) described in section \ref{sec:3b}, the term $\tilde{D}_{i,\mathbf{k}} $ that includes the Ekman and hyperviscosity damping is given by
\begin{equation}\label{eq:damp}
    \begin{aligned}
     \tilde{D}_{1,\mathbf{k}} &=  -\nu|\mathbf{k}|^{2s}\left((|\mathbf{k}|^2+k_d^2)|\mathbf{k}|^2\psi_{1,\mathbf{k}} -
    \frac{k_d^2}{2}h_{\mathbf{k}}\right)- \kappa \frac{k_d^2}{2} |\mathbf{k}|^2 \psi_{2,\mathbf{k}} ,
    \\
    \tilde{D}_{2,\mathbf{k}} &=  -\nu|\mathbf{k}|^{2s}\left((|\mathbf{k}|^2+k_d^2)|\mathbf{k}|^2\psi_{2,\mathbf{k}} - (|\mathbf{k}|^2 + \frac{k_d^2}{2}) h_{\mathbf{k}}\right) - \kappa(|\mathbf{k}|^2+\frac{k_d^2}{2}) |\mathbf{k}|^2 \psi_{2,\mathbf{k}} ,
    \end{aligned}
\end{equation}
and the Jacobian term $\tilde{J}_{i,\mathbf{k}}$ is
\begin{equation}\label{eq:jacobi}
    \small{\begin{aligned}
        \tilde{J}_{1,\mathbf{k}} &=  \sum^{\mathcal{K},\mathcal{K}}_{\substack{\mathbf{m},\mathbf{n} \\ \mathbf{m} +\mathbf{n}=\mathbf{k}}} \begin{vmatrix}
        \mathbf{m}\\
        \mathbf{n}
        \end{vmatrix} \left(-(|\mathbf{k}|^2+\frac{k_d^2}{2})(|\mathbf{m}|^2+\frac{k_d^2}{2})\psi_{1,\mathbf{n}}\psi_{1,\mathbf{m}} + \frac{k_d^2}{2}(|\mathbf{k}|^2+\frac{k_d^2}{2})\psi_{1,\mathbf{n}}\psi_{2,\mathbf{m}} +  \frac{k_d^4}{4}\psi_{2,\mathbf{n}}\psi_{1,\mathbf{m}} + \frac{k_d^2}{2}\psi_{2,\mathbf{n}} h_{\mathbf{m}} \right) ,\\
        \tilde{J}_{2 ,\mathbf{k}} &=
        \sum^{\mathcal{K},\mathcal{K}}_{\substack{\mathbf{m},\mathbf{n} \\ \mathbf{m} +\mathbf{n}=\mathbf{k}}} \begin{vmatrix}
        \mathbf{m}\\
        \mathbf{n}
        \end{vmatrix} \left(-\frac{k_d^2}{2}(|\mathbf{m}|^2+\frac{k_d^2}{2})\psi_{1,\mathbf{n}}\psi_{1,\mathbf{m}} + \frac{k_d^4}{4}\psi_{1,\mathbf{n}}\psi_{2,\mathbf{m}} +  \frac{k_d^2}{2}(|\mathbf{k}|^2+\frac{k_d^2}{2})\psi_{2,\mathbf{n}}\psi_{1,\mathbf{m}} + (|\mathbf{k}|^2+\frac{k_d^2}{2})\psi_{2,\mathbf{n}} h_{\mathbf{m}}  \right),
    \end{aligned}}
\end{equation}
in which $(m_xn_y-m_yn_x)$ is written in the form of determinant.

After introducing noises, the modified two-layer QG system (\ref{eq:cg_qg}) can be written as the conditional Gaussian stochastic model (\ref{eq:cg_2layer}) when the upper-layer flow is fully observed. The coefficients
$\mathbf{A}_0(\boldsymbol{\Psi}_1,t)$ and $\mathbf{a}_0(\boldsymbol{\Psi}_1,t)$ are vectors with the elements corresponding to wave number $\mathbf{k}$ given by
\begin{equation}\label{eq:A0a0}
    \begin{aligned}
    \mathbf{A}_{0,\mathbf{k}} =& C_{\mathbf{k}}
    \left(
         ik_x \left(
         \left((|\mathbf{k}|^2+\frac{k_d^2}{2}) \beta -         |\mathbf{k}|^4U\right)\hat{\psi}_{1,\mathbf{k}}
         - \frac{k_d^2}{2}U\hat{h}_{\mathbf{k}}
         \right)
         \right. \\
            & \left.
          -\nu|\mathbf{k}|^{2s}\left((|\mathbf{k}|^2+k_d^2)|\mathbf{k}|^2\hat{\psi}_{1,\mathbf{k}} - \frac{k_d^2}{2}\hat{h}_{\mathbf{k}}\right)
          -\sum^{\mathcal{K},\mathcal{K}}_{\substack{\mathbf{m},\mathbf{n} \\ \mathbf{m} +\mathbf{n}=\mathbf{k}}} \begin{vmatrix}
        \mathbf{m}\\
        \mathbf{n}
        \end{vmatrix} \left((|\mathbf{k}|^2+\frac{k_d^2}{2}) (|\mathbf{m}|^2+\frac{k_d^2}{2}) \hat{\psi}_{1,\mathbf{n}}\hat{\psi}_{1,\mathbf{m}}\right)
    \right) \\
\mathbf{a}_{0,\mathbf{k}} =&
        C_{\mathbf{k}}
    \left(
         ik_x \left(
         \left( \frac{k_d^2}{2}\beta - k_d^2|\mathbf{k}|^2U\right)\hat{\psi}_{1,\mathbf{k}}
         - (|\mathbf{k}|^2+\frac{k_d^2}{2})U\hat{h}_{\mathbf{k}} \right)
        \right. \\
        & \left.
         +\nu|\mathbf{k}|^{2s}(|\mathbf{k}|^2+\frac{k_d^2}{2})\hat{h}_{\mathbf{k}}
         -\sum^{\mathcal{K},\mathcal{K}}_{\substack{\mathbf{m},\mathbf{n} \\ \mathbf{m} +\mathbf{n}=\mathbf{k}}} \begin{vmatrix}
        \mathbf{m}\\
        \mathbf{n}
        \end{vmatrix} \left(\frac{k_d^2}{2}(|\mathbf{m}|^2+\frac{k_d^2}{2})\hat{\psi}_{1,\mathbf{n}}\hat{\psi}_{1,\mathbf{m}}\right)
    \right)
    \end{aligned}
\end{equation}
$\bm{\mathsf{A}}_1(\boldsymbol{\Psi}_1,t)$ and $\bm{\mathsf{a}}_1(\boldsymbol{\Psi}_1,t)$ are matrices with the entries corresponding to wave numbers $\mathbf{k}$ in row and wave number $\mathbf{m}$ in column given by
\begin{equation}\label{eq:a1A1}
    \begin{aligned}
    \bm{\mathsf{A}}_{1,(\mathbf{k},\mathbf{m)}} &= C_{\mathbf{k}}
    \left(
         ik_x (\frac{k_d^2}{2} \beta + k_d^2|\mathbf{k}|^2U - \kappa\frac{\frac{k_d^2}{2} |\mathbf{k}|^2}{ik_x})\delta_{\mathbf{k},\mathbf{m}} +
         \begin{vmatrix}
        \mathbf{m}\\
        \mathbf{k} -\mathbf{m}
        \end{vmatrix} \frac{k_d^2}{2}(|\mathbf{k}|^2\hat{\psi}_{1,\mathbf{k} -\mathbf{m} } - \hat{h}_{\mathbf{k} -\mathbf{m} })
    \right) \\
        \bm{\mathsf{a}}_{1,(\mathbf{k},\mathbf{m})} &=
        C_{\mathbf{k}}
    \left(
         ik_x
         \left((|\mathbf{k}|^2+\frac{k_d^2}{2}) \beta + |\mathbf{k}|^4U - \kappa\frac{(|\mathbf{k}|^2+\frac{k_d^2}{2}) |\mathbf{k}|^2}{ik_x}
         -\nu|\mathbf{k}|^{2s}\frac{(|\mathbf{k}|^2+k_d^2) |\mathbf{k}|^2}{ik_x} \right)\delta_{\mathbf{k},\mathbf{m}}
         \right. \\
            & \left.
         -
         \begin{vmatrix}
        \mathbf{m}\\
        \mathbf{k} -\mathbf{m}
        \end{vmatrix} \left(|\mathbf{k}|^2 \frac{k_d^2}{2}\hat{\psi}_{1,\mathbf{k}-\mathbf{m}} + (|\mathbf{k}|^2+\frac{k_d^2}{2})\hat{h}_{\mathbf{k} -\mathbf{m}}\right)
    \right)
    \end{aligned}
\end{equation}
where $\delta_{\mathbf{k}, \mathbf{m}}=1$ if $\mathbf{k}=\mathbf{m}$ else $\delta_{\mathbf{k}, \mathbf{m}}=0$.

% \subsection{Numerical results}
% \begin{table}[h]
% \begin{center}
% \begin{tabular}{cc}
% \topline
% one-step CGDA &  0.369  \\
% multi-step CGDA & 0.150 \\
% \botline
% \end{tabular}
% \end{center}
% \caption{Mean RMSEs for one-step CGDA and multi-step CGDA with upper-layer flow directly observed}\label{tab:rmse_flowobs}
% \end{table}

% \begin{figure}[h]
%  \centerline{\includegraphics[width=33pc]{rmses_flowobs_K128_beta22_tr_real.png}}
%   \caption{Spatial average RMSEs of LSM-DA and CG DA with upper layer fully observed.}\label{fig:rmses_beta22_flowobs}
% \end{figure}

%%%%%%%%%%%%%%%%%%%%%%%%%%%%%%%%%%%%%%%%%%%%%%%%%%%%%%%%%%%%%%%%%%%%%
% REFERENCES
%%%%%%%%%%%%%%%%%%%%%%%%%%%%%%%%%%%%%%%%%%%%%%%%%%%%%%%%%%%%%%%%%%%%%
\bibliographystyle{ametsocV6}
\bibliography{references}

\end{document}